\documentclass[conference,a4paper]{IEEEtran}
\usepackage{xcolor}
\usepackage{balance}

\ifCLASSINFOpdf
  \usepackage[pdftex]{graphicx}
\else
  \usepackage[dvips]{graphicx}
\fi
\usepackage{amsmath,amssymb}
\usepackage{physics}
\ifCLASSOPTIONcompsoc
 \usepackage[caption=false,font=normalsize,labelfont=sf,textfont=sf]{subfig}
\else
 \usepackage[caption=false,font=footnotesize]{subfig}
\fi
\usepackage{cite}

\usepackage[colorlinks]{hyperref}
\usepackage{xcolor}
\hypersetup{
colorlinks,
    linkcolor={red!50!black},
    citecolor={blue!50!black},
    urlcolor={blue!80!black}
}

\usepackage{mathtools}

\def\XXint#1#2#3{{\setbox0=\hbox{$#1{#2#3}{\int}$}
     \vcenter{\hbox{$#2#3$}}\kern-.5\wd0}}

\usepackage{bm}
\newcommand{\R}{\mathbb{R}}

\newcommand{\ie}{\textit{i.e.}\/, }
\newcommand{\eg}{\textit{e.g.}\/, }
\newcommand{\cf}{\textit{cf.}\/, }

\providecommand*{\mrm}[1]{\mathrm{#1}}
\providecommand*{\unit}[1]{\ensuremath{\mrm{\,#1}}}
\providecommand*{\eu}{\ensuremath{\mrm{e}}}
\providecommand*{\iu}{\ensuremath{\mrm{i}}}

\renewcommand{\Re}{\ensuremath{\mrm{Re}}}	
\renewcommand{\Im}{\ensuremath{\mrm{Im}}}	

\providecommand*{\degree}{\ensuremath{^\circ}}

\begin{document}
%
\title{On the interpretation and significance of the fluctuation-dissipation theorem in infrared spectroscopy}

\author{\IEEEauthorblockN{
Sven~Nordebo\IEEEauthorrefmark{1},   
}                                     
\IEEEauthorblockA{\IEEEauthorrefmark{1}
Department of Physics and Electrical Engineering, Linn\ae us University,   351 95 V\"{a}xj\"{o}, Sweden. E-mail: sven.nordebo@lnu.se} 
}



\maketitle

\begin{abstract}
In this paper we revisit the classical fluctuation-dissipation theorem with derivations and interpretations based on
quantum electrodynamics (QED). As a starting point we take the widely cited semiclassical expression of the theorem
connecting the absorption coefficient with the correlation spectra of a radiating molecular dipole.
The literature is suggesting how this connection can be derived in terms of quantum mechanical statistical averages,
but the corresponding results in terms of QED seems to be very difficult to trace in detail. The problem is therefore addressed here based on first principles.
Interestingly, it turns out that the QED approach applied to the aforementioned statistical averages does not only prove the validity of the
fluctuation-dissipation theorem, but it also provides a derivation and a quantum mechanical interpretation of Schwarzschild's equation for radiative transfer.
In particular, it is found that the classical Beer-Bouguer-Lambert law is due to absorption as well as of stimulated emission, 
and furthermore that the source term in Schwarzschild's equation (Kirchhoff's law) is due solely to spontaneous emission.
The significance of the fluctuation-dissipation theorem is finally elaborated on in terms of the appropriate scaling of line strength parameters (including line mixing) 
which is relevant in far infrared and millimeter wave broadband applications.
\end{abstract}

\vskip0.5\baselineskip

\section{Introduction}



The fluctuation-dissipation theorem (FDT) is a term commonly used to quantify the relationship between dissipation and correlation spectra 
and it has its roots already with Nyquist in 1928 \cite{Nyquist1928}, \cf also  \cite{Kubo1957,Kubo1966} and \cite[p.~26]{Forster1990}. 
In general terms, the dissipation refers here to the response of a given system to an external disturbance and the fluctuation refers to the internal fluctuations
of the system in the absence of the disturbance \cite{Kubo1966}. 
More precisely, the dissipative process also comprises an irreversible transformation of energy in one form (here electromagnetic) into another (here thermal), \cf \eg \cite{Kubo1957,Kubo1966}.

Later developments on this topic include works by \eg Callen and Welton \cite{Callen+Welton1951}, Kubo \cite{Kubo1957,Kubo1966}, 
Huber and Van Vleck \cite{Huber+VanVleck1966,VanVleck+Huber1977} and Davies et al \cite{Davies+etal1982}.
General aspects of fluctuation-dissipation relations in statistical physics are considered in \eg \cite{Marconi+etal2008}.
The fluctuation-dissipation theorem is also closely related to the asymmetry $I(-\omega)=\eu^{-\beta\hbar\omega}I(\omega)$ where
$I(\omega)$ is the Fourier transform of the dipole autocorrelation function (spectral density)
and $\beta=1/k_\mrm{B}T$ where $k_\mrm{B}$ is Boltzmanns constant and $T$ the temperature, \cf \cite[Eq.~(II.31)]{Hartmann+etal2021} and \cite[Eq.~(A10.12)]{Rayer+etal2020}. 
It is in fact this property that enables the presence of two different but yet equivalent formulations of 
the theorem involving the factors $\tanh(\beta\hbar\omega/2)$ and $\left(1-\eu^{-\beta\hbar\omega}\right)$, respectively, 
depending on the definition of the associated correlation spectra, \cf \cite[Eq.~(II.36)]{Hartmann+etal2021} and \cite[Eq.~(A10.14)]{Rayer+etal2020}.
In this paper we will adhere to the latter formulation.

In infrared spectroscopy the following version of the fluct-uation-dissipation theorem is oftenly referred to
\begin{equation}\label{eq:sigmaaexprGauss}
\sigma_\mrm{a}(\omega)=\frac{4\pi^2\omega}{3\hbar\mrm{c_0}}I(\omega)\left(1-\eu^{-\beta\hbar\omega}\right), 
\end{equation}
where $\sigma_\mrm{a}(\omega)$ is the absorption cross section of the radiatively active molecule, $\omega$ the angular frequency of the radiation, 
$\mrm{c}_0=1/\sqrt{\mu_0\epsilon_0}$ the speed of light in vacuum
and $I(\omega)$ the spectral density, see \eg \cite[p.~3084]{Gordon1966a}, \cite[p.~111]{Filippov+Tonkov1993}, \cite[p.~784]{Tonkov+etal1996} 
and \cite[p.~14]{Hartmann+etal2021}.
The precise form of the theorem as given by \eqref{eq:sigmaaexprGauss} is however not easy to trace in the literature.
As \eg it is not at all obvious from the given references that the formula \eqref{eq:sigmaaexprGauss} is given here in 
Gaussian units. 
It may therefore be of interest to revisit the  fluctuation-dissipation theorem and to derive the result \eqref{eq:sigmaaexprGauss}
from first principles. To this end, it is mentioned in \cite[p.~13]{Hartmann+etal2021} with a reference to \cite{Robert+Galatry1971}
that the absorption cross section can be defined based on quantum averages as
\begin{equation}\label{eq:sigmaadef}
\sigma_\mrm{a}(\omega)=\frac{P_\mrm{r}}{S_\mrm{i}}
\end{equation}
where
\begin{eqnarray}
P_\mrm{r} & = & \displaystyle\lim_{T\rightarrow\infty}\frac{1}{T}\left(\trace\{\rho(0)H_{\bm{k}}\} - \trace\{\rho(T)H_{\bm{k}}\}\right) \label{eq:Padef} \\
S_\mrm{i} & = & \frac{1}{\mu_0}\trace\left\{\rho_{\bm{k}}{\cal EB} \right\} \label{eq:Sidef}
\end{eqnarray}
and where $P_\mrm{r}$ is the power removed from a single radiation cavity mode indexed $\bm{k}$ and $S_\mrm{i}$ is the corresponding intensity.
Here, $\trace\{\cdot\}$ denotes the trace and $\rho(t)$ is the density operator satisfying the Liouville-von Neumann equation with respect to
the total molecular system including the radiation, $H_{\bm{k}}$ is the Hamiltonian of the radiation mode with density $\rho_{\bm{k}}$ and
${\cal E}$ and ${\cal B}$ are the electric field and magnetic flux operators, respectively, \cf  \cite[Chapt. II.2]{Hartmann+etal2021} and \cite{Loudon2000}.
The interaction potential connecting the electromagnetic field with the molecular system is as usual given by $-\bm{d}\cdot\bm{{\cal E}}$ where $\bm{d}$ is the dipole operator.

Very interestingly, it turns out that a quantum electrodynamical (QED) treatment of the expressions \eqref{eq:Padef} and \eqref{eq:Sidef}
within the usual second order perturbation analysis does not give directly the result \eqref{eq:sigmaaexprGauss}.
As we will show in this paper, we find instead the following presumably new and more general interpretation of \eqref{eq:Padef} and \eqref{eq:Sidef} where
\begin{eqnarray}
P_\mrm{r} & = & \frac{2\pi{\cal E}_0^2\omega^3}{3\hbar}I(\omega)\left(\langle n \rangle (1-\eu^{-\beta\hbar\omega})-\eu^{-\beta\hbar\omega} \right) \label{eq:Pares} \\
S_\mrm{i} & = & \frac{2{\cal E}_0^2\omega^2}{\eta_0} \left(\langle n \rangle+\frac{1}{2} \right)\label{eq:Sires}
\end{eqnarray}
and where $\eta_0=\sqrt{\mu_0/\epsilon_0}$ is the wave impedance of vacuum and ${\cal E}_0^2=\hbar/2\epsilon_0\omega V_\mrm{c}$ 
where $V_\mrm{c}$ is the volume of the cavity, \cf \cite[Chapt.~4]{Loudon2000}. 
Notably, in \eqref{eq:Pares} has also been included a division by 3 to account for isotropic conditions by taking an average over all polarization directions of the incident light.
Finally, $\langle n \rangle$ denotes here the mean number of photons in cavity mode $\bm{k}$ with respect to the density $\rho_{\bm{k}}$.

As we can see from \eqref{eq:Pares} and \eqref {eq:Sires}, these expressions can only define
an absorption coefficient in the sense of \eqref{eq:sigmaadef} independent of $\langle n \rangle$ when $\langle n \rangle\rightarrow \infty$,
in which case we obtain 
\begin{equation}\label{eq:sigmaaexpr}
\sigma_\mrm{a}(\omega)=\frac{\pi\eta_0\omega}{3\hbar}I(\omega)\left(1-\eu^{-\beta\hbar\omega}\right).
\end{equation}
The expression \eqref{eq:sigmaaexpr} is given here in SI-units and is equivalent to \eqref{eq:sigmaaexprGauss} when translated to Gaussian units.
Now, the assumption that $\langle n \rangle\rightarrow \infty$ will be adequate if we consider a beam of light with a well defined direction
and a continuous flow of power, such as a laser beam. However, in a more general situation with thermally induced radiative transfer in the atmosphere,
the mean value $\langle n \rangle$ will not be very large and the definitions \eqref{eq:sigmaadef}, \eqref{eq:Padef} and \eqref{eq:Sidef} will not imply \eqref{eq:sigmaaexpr}. 
As \eg based on the Planck distribution of black body radiation \cite[Eq.~(4.6.21)]{Loudon2000} we can calculate that the mean number of photons in a cavity mode is less than 1
for wavenumbers larger than 142\unit{cm^{-1}}.
In this paper, we will resolve this apparent paradox and show that there is no contradiction here provided that \eqref{eq:Padef} and \eqref{eq:Sidef} are properly interpreted. 
In particular, in the case of thermally induced radiative transfer when $\langle n \rangle$ is finite, we will show that 
Schwarzschild's equation \cite[pp.~29-30]{Liou2002} follows directly from the QED results \eqref{eq:Pares} and \eqref{eq:Sires},
and in which case the fluctuation-dissipation theorem as stated in \eqref{eq:sigmaaexpr} is an inalienable integral part.
The key observation here is that $P_\mrm{r}$ as defined in \eqref{eq:Padef} is not the {\em absorbed} power, but rather the power that is {\em removed} from the radiation.
Thus, according to the latter interpretation, $P_\mrm{r}$ also encompasses the power that is thermally re-radiated, all according to Kirchhoff's law
\cf \cite[pp.~13-14]{Liou2002} and \cite[p.~160]{Ishimaru1997}, and it is therefore consistent with Schwarzschild's equation.
To this end, an interesting interpretation of \eqref{eq:Pares} is then that the term proportional to $\langle n \rangle$  represents the absorption and the
term proportional to $\eu^{-\beta\hbar\omega}(\langle n \rangle+1)$ represents the stimulated and spontaneous emission, \cf also \cite[Eq.~(2.102)]{Sakurai1967}.

Regarding the significance of the fluctuation-dissipation theorem in infrared spectroscopy we will emphasize the following.
The line strength parameters $S_n$ are usually defined with respect to a calculation of absorption which is based solely on the spectral density $I(\omega)$
without reference to the FDT, 
\cf \eg \cite[Eqs.~(1.3.10), (1.3.11), (4.2.13) and (D.18)]{Liou2002}, \cite{Tennyson+etal2014} and \cite{HITRANdefs}.
This is certainly relevant at large wavenumber ranges, say more than 100\unit{cm^{-1}}, depending on the level of accuracy that is required.
However, at lower wavenumbers such as with millimeter waves, and in the band wings where line mixing effects are important,
it may also become important to incorporate the frequency scaling $S_n/\omega_n\left(1-\eu^{-\beta\hbar\omega_n}\right)$
in order to take the fluctuation-dissipation theorem \eqref{eq:sigmaaexpr} into proper account. 
Physically this means that the connection between fluctuation and dissipation takes place over a (relative) bandwidth 
which is sufficiently large so that the factor $\omega\left(1-\eu^{-\beta\hbar\omega}\right)$ in \eqref{eq:sigmaaexpr} will play a significant role.
This is certainly an issue that is already well known and it has been taken into consideration previously in \eg \cite[Eq.~(1)]{Rosenkranz1975}. 
However, a formal derivation of the aforementioned scaling with respect to parameters found in databases
and its consequences for isolated lines as well as for line mixing formulas seems previously 
to have been missing. This scaling was therefore introduced in \cite{Nordebo2023a}, and its significance
in the context of the fluctuation-dissipation theorem is further elaborated on and exemplified in this paper.

The rest of the paper is organized as follows. A detailed account on the interpretation and significance of the fluctuation-dissipation theorem (FDT) as outlined above is given
in sections \ref{sect:Interpretation} and \ref{eq:Significance}, respectively.
The result \eqref{eq:Pares} and the derivation of Schwarzschild's equation based on QED are given in sections \ref{sect:Dissipation} and \ref{sect:Schwarzschild}, respectively.
Some of the more lengthy details in deriving \eqref{eq:Pares} are given in Appendix \ref{sect:abspower} and the vector nature of light is briefly discussed in  Appendix \ref{sect:vector}.
A summary is finally given in section \ref{sect:Summary}.

\section{Interpretation of the fluctuation-dissipation theorem in terms of QED}\label{sect:Interpretation}

We will now derive the results \eqref{eq:Pares}, \eqref{eq:Sires} and \eqref{eq:sigmaaexpr} in detail,
their interpretation in terms of quantum electrodynamics (QED) as well as their connection to Schwarzschild's equation of radiative transfer. 
To simplify the notation the dipole and electric field operators are treated here as scalars and the implications of the vector nature of light is finally addressed in Appendix \ref{sect:vector}.

\subsection{Fluctuation} 
We start by defining the dipole autocorrelation function and briefly comment on some of its most important properties, \cf also \cite[Sect.~II.2 and II.3]{Hartmann+etal2021}.
The dipole autocorrelation function is defined in the Heisenberg picture as usual 
\begin{equation}\label{eq:QMCtsummary1}
C(t)=\trace\{\eu^{\iu H_\mrm{M}t/\hbar} \tilde{d}^\dagger \eu^{-\iu H_\mrm{M}t/\hbar}\tilde{d} \rho_\mrm{M}\},
\end{equation}
where $\tilde{d}=\eu^{\iu\bm{k}\cdot\bm{r}}d$, \cf \cite[Eq.~(II.8)]{Hartmann+etal2021}.
Here, $H_\mrm{M}$ is the Hamiltonian of the total molecular system consisting of a single absorber together 
with a neighborhood of colliding molecules which are radiatively inactive. The corresponding canonical density operator is given by 
\begin{equation}\label{eq:rhoMdef}
\rho_\mrm{M}=\frac{\eu^{-\beta H_\mrm{M}}}{\trace\{\eu^{-\beta H_\mrm{M}}\}},
\end{equation}
where $\beta=1/k_\mrm{B}T$.
The dipole and position operators of the absorbing molecule are denoted $d$ and $\bm{r}$, respectively. 
It may be noted here that even though it is not necessary for the results in this paper, we treat the position $\bm{r}$ as an operator in consistency with \cite[Eq.~ (II.9)]{Hartmann+etal2021}.
The wave vector of the incident radiation field is denoted $\bm{k}=k\hat{\bm{k}}$ where $k=\omega/\mrm{c}_0$ is the wavenumber.
The operator products in \eqref{eq:QMCtsummary1} should be understood in the sense of tensor products including 
all the degrees of freedom of all the interacting molecules (including the radiator itself), and where 
$d$ and $\bm{r}$ only represent two of the ``internal'' degrees of freedom, \cf \cite[Sect.~II.2]{Hartmann+etal2021}. 
To this end, it is noted that the (modified) dipole operator $\tilde{d}=\eu^{\iu\bm{k}\cdot\bm{r}}d$ should also be understood in the sense of a tensor product and 
hence that its two factors commute. 

The spectral density $I(\omega)$ is defined as the Fourier transform of the dipole autocorrelation function
\begin{equation}\label{eq:IomegaFT0}
I(\omega)={\cal F}\{C(t)\}=\frac{1}{2\pi}\int_{-\infty}^{\infty}C(t)\eu^{\iu\omega t}\mrm{d}t,
\end{equation}
with inverse
\begin{equation}\label{eq:IomegaIFT0}
C(t)={\cal F}^{-1}\{I(\omega)\}=\int_{-\infty}^{\infty}I(\omega)\eu^{-\iu\omega t}\mrm{d}\omega,
\end{equation}
and it is noted that $C(0)=\int_{-\infty}^{\infty}I(\omega)\mrm{d}\omega$. 

The expression \eqref{eq:QMCtsummary1} can readily be expanded in standard quantum mechanical notation as
\begin{multline}\label{eq:QMCtexpr1}
C(t)=\bra{l}\eu^{\iu H_\mrm{M}t/\hbar} \tilde{d}^\dagger\ket{l^\prime}\bra{l^\prime}\eu^{-\iu H_\mrm{M}t/\hbar}\tilde{d}\rho_\mrm{M}\ket{l} \\
=\eu^{\iu \omega_lt}\bra{l} \tilde{d}^\dagger\ket{l^\prime}\eu^{-\iu \omega_{l^\prime}t}\bra{l^\prime}\tilde{d}\ket{l}\frac{\eu^{-\beta E_l}}{\trace\{\rho_\mrm{M}\}} \\
=\eu^{-\iu \omega_{l^\prime l}t} \left|\bra{l^\prime}\tilde{d}\ket{l}\right|^2\frac{\eu^{-\beta E_l}}{\trace\{\rho_\mrm{M}\}},
\end{multline}
where we have employed Einsteins summation convention and the eigenvalue relation 
$H_\mrm{M}\ket{l}=E_l\ket{l}$ where $E_l=\hbar\omega_l$ and $\omega_{l^\prime l}=\omega_{l^\prime}-\omega_l$. 
From \eqref{eq:QMCtexpr1} it is now immediately clear
that the dipole autocorrelation function satisfies the complex conjugated symmetry $C^*(t)=C(-t)$.
Due to this symmetry, we can now evaluate the spectral density as
\begin{equation}\label{eq:ComegaFL0}
I(\omega)=\frac{1}{\pi}\Re\{\widetilde{C}(\omega)\}
\end{equation}
for all $\omega\in\R$ and where $\widetilde{C}(\omega)$ denotes the Fourier-Laplace transform
\begin{equation}\label{eq:Comega0}
\widetilde{C}(\omega)=\int_{0}^{\infty}C(t)\eu^{\iu\omega t}\mrm{d}t,
\end{equation}
with region of analyticity in an upper half plane including $\Im\{\omega\}=0$.

We will now derive the important asymmetry of the spectral density $I(-\omega)=\eu^{-\beta \hbar\omega}I(\omega)$
which is closely related to the fluctuation-dissipation theorem \cite[Eq.~(II.31)]{Hartmann+etal2021}. 
However, we will not make the usual dipole approximation here ($\eu^{\iu\bm{k}\cdot\bm{r}}\approx 1$) 
but instead keep the dependence on the wave vector $\bm{k}$ in order to better facilitate an interpretation in terms of QED. 
Hence, we extend the notation introduced in \eqref{eq:QMCtsummary1} accordingly 
and write the result \eqref{eq:QMCtexpr1} as
\begin{equation}\label{eq:QMCtexpr2}
C(\bm{k},t)=\eu^{-\iu \omega_{l^\prime l}t} \left|\bra{l^\prime}\eu^{\iu\bm{k}\cdot\bm{r}}d\ket{l}\right|^2\frac{\eu^{-\beta \hbar\omega_l}}{\trace\{\rho_\mrm{M}\}}.
\end{equation}
By taking the Fourier transform of \eqref{eq:QMCtexpr2} we obtain
\begin{equation}\label{eq:Iomegaexpr1}
I(\bm{k},\omega)= \frac{\eu^{-\beta \hbar\omega_l}}{\trace\{\rho_\mrm{M}\}}\left|\bra{l^\prime}\eu^{\iu\bm{k}\cdot\bm{r}}d\ket{l}\right|^2\delta(\omega-\omega_{l^\prime l}),
\end{equation}
where ${\cal F}\{\eu^{-\iu\omega_0 t}\}=\delta(\omega-\omega_0)$ and where $\delta(\cdot)$ denotes the Dirac delta-function.
We can then evaluate $I(-\bm{k},-\omega)$ as follows
\begin{multline}
I(-\bm{k},-\omega)= \frac{\eu^{-\beta \hbar\omega_l}}{\trace\{\rho_\mrm{M}\}}\left|\bra{l^\prime}\eu^{-\iu\bm{k}\cdot\bm{r}}d\ket{l}\right|^2\delta(-\omega-\omega_{l^\prime l}) \\
=\frac{\eu^{-\beta \hbar\omega_l}}{\trace\{\rho_\mrm{M}\}}\left|\bra{l^\prime}\eu^{-\iu\bm{k}\cdot\bm{r}}d\ket{l}\right|^2\delta(\omega-\omega_{ll^\prime}) \\
=\frac{\eu^{-\beta \hbar\omega_{l^\prime}}}{\trace\{\rho_\mrm{M}\}}\left|\bra{l}\eu^{-\iu\bm{k}\cdot\bm{r}}d\ket{l^\prime}\right|^2\delta(\omega-\omega_{l^\prime l}) \\
=\eu^{-\beta\hbar\omega_{l^\prime l}}\frac{\eu^{-\beta \hbar\omega_{l}}}{\trace\{\rho_\mrm{M}\}}\left|\bra{l}\eu^{-\iu\bm{k}\cdot\bm{r}}d\ket{l^\prime}\right|^2\delta(\omega-\omega_{l^\prime l}) \\
=\eu^{-\beta\hbar\omega}\frac{\eu^{-\beta \hbar\omega_{l}}}{\trace\{\rho_\mrm{M}\}}\left|\bra{l}\eu^{-\iu\bm{k}\cdot\bm{r}}d\ket{l^\prime}\right|^2\delta(\omega-\omega_{l^\prime l}),
\end{multline}
and where the last line gives the result
\begin{equation}\label{eq:fluctdisstheor2}
I(-\bm{k},-\omega)=\eu^{-\beta\hbar\omega}I(\bm{k},\omega).
\end{equation}
Apparently, we can interpret this asymmetry as a consequence of dipolar fluctuations in thermal equilibrium.

\subsection{Dissipation}\label{sect:Dissipation}

\subsubsection{Problem setup}
The dissipation part of the FDT is due to the electromagnetic disturbance of the system in thermal equilibrium.
The time evolution of this disturbance is governed by the Schrödinger equation
\begin{equation}\label{eq:Schrodingerequation3}
\iu\hbar\frac{\partial}{\partial t}\ket{\psi(t)}=H(t)\ket{\psi(t)},
\end{equation}
for $t\geq 0$, and where the Hamiltonian is given by
\begin{equation}\label{eq:Hoft}
H(t)=H_\mrm{M}+H_{\bm{k}}-d{\cal E},
\end{equation}
where $H_\mrm{M}$ is the Hamiltonian of the total molecular system as described above.
Here, $H_{\bm{k}}$  is furthermore denoting the Hamiltonian of a radiation ``cavity-mode'' associated with the wave vector $\bm{k}$ where $k=|\bm{k}|=\omega/\mrm{c}_0$,
and $-d{\cal E}$ is the interaction potential where $d$ is the dipole operator of the absorbing molecule and ${\cal E}$ the electric field operator, \cite[Eq.~(4.8.31)]{Loudon2000}.
The radiation Hamiltonian is that of a harmonic oscillator associated with number states $\ket{n}$ where
\begin{equation}
H_{\bm{k}}\ket{n}=\left(n+\frac{1}{2}\right)\hbar\omega\ket{n},
\end{equation}
for $n=0,1,2,\ldots$. The electric field and magnetic flux operators are given here as
\begin{equation}\label{eq:EB}
\left\{\begin{array}{l}
{\cal E}={\cal E}_0\iu\omega\left(\hat{a}\eu^{\iu \bm{k}\cdot\bm{r}}-\hat{a}^\dagger\eu^{-\iu \bm{k}\cdot\bm{r}} \right), \vspace{0.2cm} \\
{\cal B}={\cal E}_0\iu k \left(\hat{a}\eu^{\iu \bm{k}\cdot\bm{r}}-\hat{a}^\dagger\eu^{-\iu \bm{k}\cdot\bm{r}} \right),
\end{array}\right.
\end{equation}
where $\hat{a}$ and $\hat{a}^\dagger$ are the usual annihilation and creation operators, respectively,
and ${\cal E}_0=\sqrt{\hbar/2\epsilon_0V_\mrm{c}\omega}$ where $V_\mrm{c}$ is the volume of the cavity, see \cite[pp.~141-142]{Loudon2000}, 
\cite[p.~319]{Grynberg+etal2010} or \cite[Chapt.~2]{Sakurai1967}. 
For simplicity, we suppress here the wavenumber and polarization indices ($\bm{k}$ and $\lambda=1,2$) of the operators associated with each cavity mode
as well as its time-dependency $\eu^{\mp\iu\omega t}$, \cf \cite[Chapt.~4.4]{Loudon2000}.
Note however that $\eu^{\pm\iu \bm{k}\cdot\bm{r}}$ are treated as operators here, $\bm{r}$ being the position operator of the absorbing molecule.
It is readily seen that both ${\cal E}$ and ${\cal B}$ are self-adjoint, \ie ${\cal E}^\dagger={\cal E}$ and ${\cal B}^\dagger={\cal B}$.
It is noted that $\hat{a}$ is associated with the annihilation of a photon
propagating in the $\bm{k}$-direction of a classical wave $\eu^{\iu \bm{k}\cdot\bm{r}}\eu^{-\iu\omega t}$, 
and that $\hat{a}^\dagger$ is associated with the creation of a photon
propagating in the same $\bm{k}$-direction of the same classical wave $\eu^{-\iu \bm{k}\cdot\bm{r}}\eu^{\iu\omega t}$.
The annihilation and creation operators satisfy the following ladder properties
\begin{equation}\label{eq:anahatn}
\left\{\begin{array}{l}
\hat{a}\ket{n}=\sqrt{n}\ket{n-1}, \vspace{0.2cm} \\
\hat{a}^\dagger\ket{n}=\sqrt{n+1}\ket{n+1},
\end{array}\right.
\end{equation}
for $n=0,1,2,\ldots$ and where $\hat{a}\ket{0}=0$. The number operator is furthermore given by $N=\hat{a}^\dagger\hat{a}$ where $N\ket{n}=n\ket{n}$.

The Hamiltonian written in \eqref{eq:Hoft} should be understood in the sense of tensor product spaces where the only interaction between the
molecular and radiation degrees of freedom is obtained via the interaction potential $-d{\cal E}$, \cf \cite[Eq.~(4.8.30)]{Loudon2000}.
The Hamiltonian $H(t)$ given by \eqref{eq:Hoft} is furthermore considered to be time-dependent in the sense that the interaction $-d{\cal E}$ is switched on at time $t=0$.
We can then invoke the theory of time-dependent perturbations in the interaction picture as described in \cite[Chapts.~5.5-5.7]{Sakurai+Napolitano2011} 
where $H=H_0+V$, $H_0=H_\mrm{M}+H_{\bm{k}}$ and $V=-d{\cal E}$.
The interaction potential written in the interaction picture is then given by
\begin{equation}\label{eq:VIexpr}
V_\mrm{I}(t)=\eu^{\iu H_0t/\hbar}V\eu^{-\iu H_0t/\hbar},
\end{equation}
and where it is readily seen that $V_\mrm{I}^\dagger(t)=V_\mrm{I}(t)$. 
The corresponding time-evolution operator $U_\mrm{I}(t)$ is defined by the Schrödinger equation
\begin{equation}\label{eq:SchrUIt}
\iu\hbar\frac{\partial}{\partial t}U_\mrm{I}(t)=V_\mrm{I}(t)U_\mrm{I}(t),
\end{equation}
together with the initial condition $U_\mrm{I}(0)=1$, and which hence satisfies the following integral equation
\begin{equation}\label{eq:IntEqUIt}
U_\mrm{I}(t)=1-\frac{\iu}{\hbar}\int_0^t V_\mrm{I}(t^\prime)U_\mrm{I}(t^\prime)\mrm{d}t^\prime.
\end{equation}
Up to second order in the interaction potential, the time-evolution operator $U_\mrm{I}(t)$ can now be written as the following truncated Dyson series
\begin{equation}\label{eq:UI2ndorder}
U_\mrm{I}(t)\approx 
1-\frac{\iu}{\hbar}\int_0^t V_\mrm{I}(t^\prime)\mrm{d}t^\prime-\frac{1}{\hbar^2} 
\int_0^t \int_0^{t^\prime}V_\mrm{I}(t^\prime)V_\mrm{I}(t^{\prime\prime})\mrm{d}t^{\prime\prime}\mrm{d}t^\prime,
\end{equation}
\cf \cite[Eq.~(5.7.6)]{Sakurai+Napolitano2011}.
It is important here to observe the order of integration in the double integral above where the two operators 
$V_\mrm{I}(t^\prime)$ and $V_\mrm{I}(t^{\prime\prime})$ are in general non-commuting.

We will now associate with \eqref{eq:Schrodingerequation3} a density operator $\rho(t)$ satisfying the following Liouville-von Neumann equation
\begin{equation}\label{eq:LiouvillevonNeumann}
\iu\hbar \frac{\partial}{\partial t}\rho(t)=\left[H(t),\rho(t) \right],
\end{equation}
\cf \cite[Eq.~(3.4.29)]{Sakurai+Napolitano2011} and \cite[Eq.~(II.3)]{Hartmann+etal2021}.
Based on the definitions made above it can now readily be verified that the solution to \eqref{eq:LiouvillevonNeumann} is given by
\begin{equation}\label{eq:rhotsol}
\rho(t)=\eu^{-\iu H_0t/\hbar}U_\mrm{I}(t)\rho(0)U_\mrm{I}^\dagger(t)\eu^{\iu H_0t/\hbar}.
\end{equation}
It is assumed that the initial density $\rho(0)$ at time $t=0$ can be written as the tensor product
\begin{equation}
\rho(0)=\rho_\mrm{M}\rho_{\bm{k}},
\end{equation}
where $\rho_\mrm{M}$ is the canonical density of the total molecular system in thermal equilibrium as given by \eqref{eq:rhoMdef}, \cf \cite[p.~14]{Hartmann+etal2021}.
The density operator $\rho_{\bm{k}}$ of the single cavity mode can be written quite generally in terms of its number states as
\begin{equation}\label{eq:rhokdef}
\rho_{\bm{k}}=w_n\ket{n}\bra{n},
\end{equation}
where Einsteins summation convention is employed and where $w_n\geq 0$ and $\sum w_n=1$.
Based on their definition as tensor products it is noticed that the two operator pairs $\rho_\mrm{M}$ and $\rho_{\bm{k}}$, as well as $\rho_\mrm{M}$ and $H_{\bm{k}}$ are commuting. 
Based on \eqref{eq:rhokdef} it can furthermore be seen that $\rho_{\bm{k}}$ and $H_{\bm{k}}$ are commuting.
Finally, and most importantly here, it follows that also $\rho(0)$ and $H_{\bm{k}}$ are commuting.

\subsubsection{Results}
It is now straightforward to calculate the mean value of the Poynting power flow in the $\bm{k}$-direction of a single cavity mode.
We will employ \eqref{eq:EB} and \eqref{eq:anahatn} and start by computing
\begin{multline}
\left(\hat{a}\eu^{\iu \bm{k}\cdot\bm{r}}-\hat{a}^\dagger\eu^{-\iu \bm{k}\cdot\bm{r}} \right)\ket{n} \\
=\sqrt{n}\ket{n-1}\eu^{\iu \bm{k}\cdot\bm{r}}-\sqrt{n+1}\ket{n+1}\eu^{-\iu \bm{k}\cdot\bm{r}}
\end{multline}
and hence by taking the adjoint and a sign-shift
\begin{multline}
\bra{n}\left(\hat{a}\eu^{\iu \bm{k}\cdot\bm{r}}-\hat{a}^\dagger\eu^{-\iu \bm{k}\cdot\bm{r}} \right) \\
=-\sqrt{n}\bra{n-1}\eu^{-\iu \bm{k}\cdot\bm{r}}+\sqrt{n+1}\bra{n+1}\eu^{\iu \bm{k}\cdot\bm{r}}.
\end{multline}
The mean Poynting power flow can now be evaluated as 
\begin{multline}\label{eq:AppAbscoeff_trEBrhok}
\frac{1}{\mu_0}\trace\left\{\rho_{\bm{k}}{\cal E}{\cal B}\right\}
=\frac{1}{\mu_0}\bra{n}{\rho_{\bm{k}}{\cal EB}}\ket{n}=\frac{1}{\mu_0}w_n\bra{n}{{\cal EB}}\ket{n} \\
=\frac{1}{\mu_0}w_n
{\cal E}_0\iu\omega {\cal E}_0\iu k
\bra{n}\left(\hat{a}\eu^{\iu \bm{k}\cdot\bm{r}}-\hat{a}^\dagger\eu^{-\iu \bm{k}\cdot\bm{r}} \right) \\
\cdot  \left(\hat{a}\eu^{\iu \bm{k}\cdot\bm{r}}-\hat{a}^\dagger\eu^{-\iu \bm{k}\cdot\bm{r}} \right)\ket{n} \\
=-\frac{{\cal E}_0^2\omega k}{\mu_0}w_n\left(-n-(n+1) \right)=\frac{2{\cal E}_0^2\omega^2}{\eta_0}w_n\left(n+1/2 \right),
\end{multline}
and which finally gives \eqref{eq:Sires} where the mean number of photons is
\begin{equation}\label{eq:sumwnn}
\langle n \rangle=\trace\{\rho_{\bm{k}}N\}=\sum_n w_nn.
\end{equation}

Based on the definitions and the results that have been elaborated on above it is now possible to show similarly that
\begin{multline}\label{eq:Pares1}
P_\mrm{r}=\displaystyle\lim_{T\rightarrow\infty}\frac{1}{T}\left(\trace\{\rho(0)H_{\bm{k}}\} - \trace\{\rho(T)H_{\bm{k}}\}\right) \\
=\frac{2\pi{\cal E}_0^2\omega^3}{3\hbar} \left(I(\bm{k},\omega)\langle n \rangle- I(-\bm{k},-\omega)(\langle n \rangle +1)\right),
\end{multline}
and where a division by 3 has been included to take the vector nature of light into account, \cf Appendix \ref{sect:vector}.
The derivation of \eqref{eq:Pares1} is straightforward but rather lengthy and is therefore presented in Appendix \ref{sect:abspower}.
By furthermore invoking the asymmetry \eqref{eq:fluctdisstheor2} due to thermal fluctuations we arrive at
\begin{multline}\label{eq:Pares2}
P_\mrm{r}
=\frac{2\pi{\cal E}_0^2\omega^3}{3\hbar} I(\bm{k},\omega)\left(\langle n \rangle- \eu^{-\beta\hbar\omega}(\langle n \rangle +1)\right),
\end{multline}
and which is the result presented in \eqref{eq:Pares}. 
An important interpretation of \eqref{eq:Pares1} and \eqref{eq:Pares2} is that 
the first term $I(\bm{k},\omega)\langle n \rangle$ is associated with the absorption (annihilation) of photons and 
$I(-\bm{k},-\omega)(\langle n \rangle +1)$ with the emission (creation) of photons, \cf also \cite[Eq.~(2.102)]{Sakurai1967}. 
And furthermore, in the latter case the term $I(-\bm{k},-\omega)\langle n \rangle$ corresponds to the stimulated emission whereas the final term $I(-\bm{k},-\omega)\cdot 1$ 
represents the spontaneous emission (emission even when $\langle n \rangle=0$). 
As we will see next, even the spontaneous emission which oftenly can be neglected, is indeed crucial here for the derivation of Schwarzschild's equation based on QED.
The fluctuation-dissipation theorem as stated in \eqref{eq:sigmaaexpr} is furthermore an inalienable integral part of this formulation.
Finally, once these interpretations are in place and under the dipole approximation ($\eu^{\iu \bm{k}\cdot\bm{r}}\approx 1$),
we are then ready to drop the $\bm{k}$-dependency and write the correlation spectra as  $I(\omega)$ instead of $I(\bm{k},\omega)$.

\subsection{Schwarzschild's equation}\label{sect:Schwarzschild}

We will now show that the QED results \eqref{eq:Pares} and \eqref{eq:Sires} leads directly to the Schwarzschild's equation of radiative transfer.
We recall that the expression \eqref{eq:Pares} gives the power $P_\mrm{r}$ that is removed from a single radiation cavity mode
due to the interaction with a single molecular dipole. The first observation here is that the condition $P_\mrm{r}=0$ implies that
\begin{equation}\label{eq:Blackbodymeann}
\langle n \rangle=\frac{\eu^{-\beta\hbar\omega}}{1-\eu^{-\beta\hbar\omega}},
\end{equation}
which is precisely the mean number of photons in thermal equilibrium according to Planck's law,  \cf \cite[Eq.~(1.3.7) and (4.6.21)]{Loudon2000}.
The corresponding radiance (\unit{W/m^2/sr/s^{-1}}) is given by the familiar Plack function of blackbody radiation  
\begin{equation}\label{eq:Planckfun}
B_\omega(\omega)=\frac{\hbar\omega^3}{4\pi^3\mrm{c}_0^2}\frac{\eu^{-\beta\hbar\omega}}{1-\eu^{-\beta\hbar\omega}},
\end{equation}
and where $\beta=1/k_\mrm{B}T$, \cf \eg \cite[Eq.~(1.2.3)]{Liou2002}.
In particular, in the derivation of \eqref{eq:Planckfun} it is first assed that the number of cavity modes $N_\omega$ in a volume $V_\mrm{c}$ and frequency interval $\mrm{d}\omega$ is given by 
\begin{equation}\label{eq:Nomegadef}
N_\omega=\rho_\omega V_\mrm{c}\mrm{d}\omega,
\end{equation}
where the modal density $\rho_\omega$ is given by
\begin{equation}\label{eq:rhoomegadef}
\rho_\omega=\frac{\omega^2}{\pi^2\mrm{c}_0^3},
\end{equation}
\cf \cite[Eq.~(1.1.10)]{Loudon2000}. The corresponding energy density is $W=\langle n \rangle \hbar\omega \rho_\omega$ and when
$\langle n \rangle$ is given by \eqref{eq:Blackbodymeann} we obtain the Planck function of isotropic blackbody radiation as $B_\omega(\omega)=W\mrm{c}_0/4\pi$, \cf 
\cite[(1.3.8)]{Loudon2000}.

More generally, we define now the radiance $I_\omega(\bm{r},\hat{\bm{s}})$ by the relation
\begin{equation}\label{eq:Iomegadef}
I_\omega(\bm{r},\hat{\bm{s}})\mrm{d}\Omega\mrm{d}\omega=\frac{2{\cal E}_0^2\omega^2}{\eta_0} \langle n \rangle,
\end{equation}
which is based on \eqref{eq:Sires} with the vacuum energy removed.
Here $\bm{r}$ refers to position (not an operator), $\hat{\bm{s}}$ to direction, $\mrm{d}\Omega$ to a differential solid angle and
$\mrm{d}\omega$ to the frequency interval. It is assumed here that $\langle n \rangle$ can be regarded to be fixed within the small bundle of light within solid angle $\mrm{d}\Omega$.
Considering a small cylindrical volume element extended in the $\hat{\bm{s}}$-direction with length $\mrm{d}s$ and cross-sectional area $\mrm{d}a$,
we can now make the following interpretation of \eqref{eq:Pares} where 
\begin{multline}\label{eq:dIomegadef}
\mrm{d}I_\omega(\bm{r},\hat{\bm{s}})\mrm{d}a\mrm{d}\Omega\mrm{d}\omega=-P_\mrm{r}N\mrm{d}a\mrm{d}s \\
=-\frac{2\pi{\cal E}_0^2\omega^3}{3\hbar}I(\omega)\left(\langle n \rangle (1-\eu^{-\beta\hbar\omega})-\eu^{-\beta\hbar\omega} \right)N\mrm{d}a\mrm{d}s,
\end{multline}
where the change in radiance is due to the negative of the removed power
and $N$ is the number density of radiatively active molecules inside the small cylinder with volume $\mrm{d}a\mrm{d}s$.
Let us now step by step rewrite this relation in order to derive the equation of transfer. Thus, by first removing $\mrm{d}a$ we can rewrite the differential as
\begin{multline}\label{eq:dIomegaexpr1}
\mrm{d}I_\omega(\bm{r},\hat{\bm{s}})\mrm{d}\Omega\mrm{d}\omega
=-\frac{2\pi{\cal E}_0^2\omega^3}{3\hbar}I(\omega)\langle n \rangle \left(1-\eu^{-\beta\hbar\omega}\right)N\mrm{d}s \\
+\frac{2\pi{\cal E}_0^2\omega^3}{3\hbar}I(\omega)\eu^{-\beta\hbar\omega}N\mrm{d}s.
\end{multline}
The first term on the right hand side of \eqref{eq:dIomegaexpr1} can now be rearranged as
\begin{multline}\label{eq:dIomegaexpr2}
-\frac{2\pi{\cal E}_0^2\omega^3}{3\hbar}I(\omega)\langle n \rangle \left(1-\eu^{-\beta\hbar\omega}\right)N\mrm{d}s \\
=-N \frac{\pi\eta_0\omega}{3\hbar}I(\omega)\left(1-\eu^{-\beta\hbar\omega}\right) \frac{2{\cal E}_0^2\omega^2}{\eta_0} \langle n \rangle\mrm{d}s \\
=-N\sigma_\mrm{a}(\omega)I_\omega(\bm{r},\hat{\bm{s}})\mrm{d}\Omega\mrm{d}\omega\mrm{d}s,
\end{multline}
and where \eqref{eq:sigmaaexpr} and \eqref{eq:Iomegadef} have been inserted in the last step.

In order to interpret the second term on the right hand side of \eqref{eq:dIomegaexpr1} we may employ \eqref{eq:Nomegadef} and choose
\begin{equation}\label{eq:dOmegadef}
\mrm{d}\Omega=\frac{4\pi}{N_\omega}=\frac{4\pi}{\rho_\omega V_\mrm{c}\mrm{d}\omega},
\end{equation}
where each individual cavity mode is associated with the same differential solid angle $\mrm{d}\Omega$ and $\int\mrm{d}\Omega=4\pi$.
Obviously, we must also assume here that $V_\mrm{c}\mrm{d}\omega\rightarrow\infty$ as $V_\mrm{c}\rightarrow\infty$ and $\mrm{d}\omega\rightarrow 0$.
By combining \eqref{eq:rhoomegadef} and \eqref{eq:dOmegadef} we obtain the following very useful relationship
\begin{equation}\label{eq:dOmegadomega}
\mrm{d}\Omega\mrm{d}\omega=\frac{4\pi^3\mrm{c}_0^3}{V_\mrm{c}\omega^2}.
\end{equation}
As \eg based on \eqref{eq:dOmegadomega} and \eqref{eq:Iomegadef} we can now see that the radiance $I_\omega(\bm{r},\hat{\bm{s}})$ can
in general be expressed as
\begin{equation}\label{eq:Iomegaradexpr}
I_\omega(\bm{r},\hat{\bm{s}})=\frac{\hbar\omega^3}{4\pi^3\mrm{c}_0^2}\langle n \rangle,
\end{equation}
and where \eqref{eq:Planckfun} is the special case with blackbody radiation.

The second term on the right hand side of \eqref{eq:dIomegaexpr1} can now be rearranged as
\begin{multline}\label{eq:dIomegaexpr3}
\frac{2\pi{\cal E}_0^2\omega^3}{3\hbar}I(\omega)\eu^{-\beta\hbar\omega}N\mrm{d}s \\
=\frac{2\pi\omega^3}{3\hbar}\frac{\hbar}{2\epsilon_0\omega V_\mrm{c}}I(\omega)\eu^{-\beta\hbar\omega}N\mrm{d}s \\
=N \frac{\pi\eta_0\omega}{3\hbar}I(\omega)\left(1-\eu^{-\beta\hbar\omega}\right) \\
\cdot \frac{\hbar\omega^3}{4\pi^3\mrm{c}_0^2}\frac{\eu^{-\beta\hbar\omega}}{1-\eu^{-\beta\hbar\omega}}
\cdot \frac{4\pi^3\mrm{c}_0^3}{V_\mrm{c}\omega^2}\mrm{d}s\\
=N \sigma_\mrm{a}(\omega)B_\omega(\omega)\mrm{d}\Omega\mrm{d}\omega\mrm{d}s
\end{multline}
where ${\cal E}_0^2=\hbar/2\epsilon_0\omega V_\mrm{c}$ and $\epsilon_0=1/\eta_0\mrm{c}_0$ and where
\eqref{eq:sigmaaexpr}, \eqref{eq:Planckfun} and \eqref{eq:dOmegadomega} have been used in the last line.
Now, by combining \eqref{eq:dIomegaexpr1}, \eqref{eq:dIomegaexpr2} and \eqref{eq:dIomegaexpr3} we arrive at
\begin{equation}\label{eq:SchwarzschildsEq}
\mrm{d}I_\omega(\bm{r},\hat{\bm{s}})=-N\sigma_\mrm{a}(\omega)I_\omega(\bm{r},\hat{\bm{s}})\mrm{d}s
+N \sigma_\mrm{a}(\omega)B_\omega(\omega)\mrm{d}s,
\end{equation}
which is Schwarzschild's equation, \cf \eg \cite[pp.~29-30)]{Liou2002}.

\subsubsection{Interpretation in terms of QED}

In the classical interpretation of Schwarzschild's equation the first term on the right hand side of \eqref{eq:SchwarzschildsEq} corresponds to the absorption of radiation
and is referred to as the Beer-Bouguer-Lambert law \cite[pp.~28-29]{Liou2002}. The second (source) term is then attributed 
to the Kirchhoff's law of thermal radiation which is taking into account the thermal re-radiation of absorbed power in proportionality to the absorption coefficient $\sigma_\mrm{a}$ and
the Planck function of blackbody radiation $B_\omega(\omega)$, \cf \eg \cite[pp.~13-14]{Liou2002} and \cite[p.~160]{Ishimaru1997}.
In particular, Kirchhoff's law also states that the absorptivity $1-\eu^{-N\sigma_\mrm{a}\mrm{d}s}$ is equal to the emissivity, all in accordance with Schwarzschild's equation.

In contrast, in the QED interpretation of Schwarzschild's equation based on \eqref{eq:Pares} and \eqref{eq:dIomegaexpr1} above, 
both terms on the right hand side of \eqref{eq:SchwarzschildsEq} are associated with emission.
In particular, with regard to the first term in \eqref{eq:SchwarzschildsEq} the fluctuation-dissipation theorem as stated in \eqref{eq:sigmaaexpr} 
is expressing a balance between the absorption and the {\em stimulated emission}, whereas the second (source) term in \eqref{eq:SchwarzschildsEq} 
is due solely to the {\em spontaneous emission} independent of $\langle n \rangle$. 
Hence, by rewriting the spontaneous emission \eqref{eq:dIomegaexpr3} in terms of the absorption coefficient $\sigma_\mrm{a}(\omega)$ given by \eqref{eq:sigmaaexpr} 
and the Planck function $B_\omega(\omega)$ given by \eqref{eq:Planckfun}, we have used QED here to prove the
validity of Kirchhoff's law in the context of Schwarzschild's equation as in \eqref{eq:SchwarzschildsEq}.
An interesting interpretation here is that without spontaneous emission as predicted by QED, there would be no source term
in Schwarzschild's equation and no transfer of thermal radiation in accordance with Kirchhoff's law.

It is noted that even in the case when $\langle n \rangle$ is very large (such as with a laser beam or a ray of sun light) and the source term in \eqref{eq:SchwarzschildsEq} can be neglected,
the QED interpretation of the fluctuation-dissipation theorem \eqref{eq:sigmaaexpr} still prescribes the presence of both absorption as well as of (stimulated) emission.
Even though if it may strike us as a contradiction to speak about an absorption coefficient in this case, there should be no confusion here. The absorption coefficient should then simply be interpreted
as a coefficient which is inherently compensating for the absorbed power that is being thermally re-radiated via stimulated emission.

The essence of this discussion is the following: Even though the classical and the QED interpretations of the Sch-warzschild's equation are different in terms of the absorption and emission 
processes as discussed above,
there is no contradiction with regard to what is physically observed in terms of radiances. Let us take the QED view and consider 
that the quantity $\langle n \rangle$ is representing the incident field. Now, if $\langle n \rangle$ is given by \eqref{eq:Blackbodymeann} 
in thermal equilibrium then $I_\omega(\bm{r},\hat{\bm{s}})=B_\omega(\omega)$ and $\mrm{d}I_\omega(\bm{r},\hat{\bm{s}})=0$.
If, on the other hand $\langle n \rangle=0$ (radiation from a very cold surface at temperature close to zero) then $I_\omega(\bm{r},\hat{\bm{s}})=0$
and $\mrm{d}I_\omega(\bm{r},\hat{\bm{s}})=N \sigma_\mrm{a}(\omega)B_\omega(\omega)\mrm{d}s$.
These two extreme examples are in full consistency with the classical interpretation of Schwarzschild's equation and Kirchhoff's law. 
And if $\langle n \rangle$ is very large, then the first term on the right hand side of \eqref{eq:SchwarzschildsEq} will dominate 
and the second term can be neglected, all in accordance with the classical  Beer-Bouguer-Lambert law.
However, in general, for the two interpretations to be equivalent in this comparison the absorption coefficient must be treated 
in accordance with the fluctuation-dissipation theorem as stated in \eqref{eq:sigmaaexpr}.

\section{Significance of the fluctuation-dissipation theorem}\label{eq:Significance}

\subsection{Parameter scalings}

A short elaboration on the parameters scalings associated with the fluctuation-dissipation theorem
which are necessary for the purpose of this exposition is given below, \cf also \cite[Sect.~4]{Nordebo2023a}. 
The wavenumber domain is introduced here by using $\omega=2\pi\mrm{c}_0\nu$ where $\nu=\lambda^{-1}$ is the wavenumber 
and $\lambda$ the wavelength of the radiation. 
The absorption coefficient \eqref{eq:sigmaaexpr}  is then given by
\begin{equation}\label{eq:sigmaaexprnu}
\sigma_\mrm{a}(\nu)=\frac{\pi\eta_0}{3\hbar}\nu\left(1-\eu^{-\beta h\mrm{c}_0\nu}\right)I(\nu),
\end{equation}
where $I(\nu)=\frac{1}{\pi}\Re\{\widetilde{C}(\nu)\}=2\pi\mrm{c}_0 I(\omega)$ and $\widetilde{C}(\nu)=2\pi\mrm{c}_0\widetilde{C}(\omega)$.
To start with we may assume that all quantities are given in SI-units so that
the factor $\pi\eta_0/3\hbar$ is given in $A^{-2}s^{-2}$, $\widetilde{C}(\nu)$ in $A^2s^2m^3$ and $\sigma_\mrm{a}$ in \unit{m^2}.
The simple sum of Lorentzian lines is then given by
\begin{equation}\label{eq:SumLor}
\widetilde{C}(\nu)=\sum_n \frac{S_n}{\gamma_{0n}^\prime-\iu(\nu-\nu_{0n}-\delta_{0n}^\prime)},
\end{equation}
where the line widths $\gamma_{0n}^\prime$, the transition wavenumbers $\nu_{0n}$ and the shifts $\delta_{0n}^\prime$ are in \unit{m^{-1}}
and the line strengths $S_n$ are in \unit{A^2s^2m^2}. It is noticed here that the line strengths can be interpreted as $S_n=\rho_n\mu_n^2$ where
$\rho_n$ is the Boltzmann factor and $\mu_n$ the transition dipole moment of each line $n$, 
\cf \cite{Filippov+Tonkov1993,Tonkov+Filippov2003,Ciurylo+Pine2000,Nordebo2023a}.

Now, the following parameter scalings are considered 
\begin{eqnarray}
S_n^\prime & = & \frac{\pi\eta_0}{3\hbar}\nu_n(1-\eu^{-\beta h\mrm{c}_0\nu_n})S_n, \label{eq:Snprimedef}\\
S_n^{\prime\prime} & = & \frac{\pi\eta_0}{3\hbar}S_n=\frac{S_n^\prime}{\nu_n(1-\eu^{-\beta h\mrm{c}_0\nu_n})}, \label{eq:Snbisdef}
\end{eqnarray}
where $S_n^\prime$ is in \unit{m} and $S_n^{\prime\prime}$ is \unit{m^2} and $\nu_n=\nu_{0n}+\delta_{0n}^\prime$.
The absorption coefficient corresponding to the Lorentzian lines, here abbreviated as Lor, can then be expressed as
\begin{equation}\label{eq:sigmaaLor}
\sigma_\mrm{a}(\nu)=\frac{1}{\pi}\Re\left\{\sum_n \frac{S_n^{\prime}}{\gamma_{0n}^\prime-\iu(\nu-\nu_{0n}-\delta_{0n}^\prime)}\right\},
\end{equation}
which is customary with \eg HITRAN\footnote{In HITRAN, the line strengths $S_n^\prime$ are given in \unit{cm}.} 
parameters \cite[Eq.~(8)--(10)]{HITRANdefs} in high-frequency and narrow band applications.
Here, the frequency dependency of the factor $\frac{\pi\eta_0}{3\hbar}\nu(1-\eu^{-\beta h\mrm{c}_0\nu})$ has been neglected and absorbed in 
the coefficient $S_n^\prime$, term by term, as defined in \eqref{eq:Snprimedef}.
Alternatively, we may take the fluctuation-dissipation theorem into full account, here abbreviated as Lor (fd), by writing \eqref{eq:sigmaaexprnu} as
\begin{equation}\label{eq:sigmaaLorFD}
\sigma_\mrm{a}(\nu)=\nu(1-\eu^{-\beta h\mrm{c}_0\nu})
\frac{1}{\pi}\Re\left\{\sum_n \frac{S_n^{\prime\prime}}{\gamma_{0n}^\prime-\iu(\nu-\nu_{0n}-\delta_{0n}^\prime)}\right\},
\end{equation}
where $S_n^{\prime\prime}$ has been defined in \eqref{eq:Snbisdef}.

When using the basic strong collision line mixing method, here abbreviated as SC, by Bulanin, Dokuchaev, Tonkov and Filippov
\cite{Dokuchaev+etal1982,Bulanin+etal1984,Filippov+Tonkov1993,Tonkov+etal1996,Filippov+etal2002,Tonkov+Filippov2003}
we can similarly express the absorption coefficient traditionally as 
\begin{equation}\label{eq:sigmaaSC}
\sigma_\mrm{a}(\nu)=\frac{1}{\pi}\Re\left\{\frac{\widetilde{C}_1^\prime(\nu)}{1-\frac{v_\mrm{s}^\prime}{C^\prime(0)}\widetilde{C}_1^\prime(\nu)} \right\}
\end{equation}
where
\begin{equation}\label{eq:Ctilde1primeSC}
\widetilde{C}_1^\prime(\nu)
=\sum_n \frac{S_n^\prime}{v_\mrm{s}^\prime-\iu\left(\nu-\nu_{0n}\right)},
\end{equation}
and 
\begin{equation}\label{eq:vssol2nu}
v_\mrm{s}^\prime=\frac{\displaystyle\sum_n \gamma_{0n}^\prime S_n^\prime}{\displaystyle\sum_n S_n^\prime},
\end{equation}
and where $C^{\prime}(0)=\sum_n S_n^{\prime}$.
More rigorously, we may also incorporate the fluctuation-dissipation theorem, here abbreviated as SC (fd), by writing
\begin{equation}\label{eq:sigmaaSCfd}
\sigma_\mrm{a}(\nu)=\nu\left(1-\eu^{-\beta h\mrm{c}_0\nu}\right)\frac{1}{\pi}\Re\left\{\frac{\widetilde{C}_1^{\prime\prime}(\nu)}
{1-\frac{v_\mrm{s}^\prime}{C^{\prime\prime}(0)}\widetilde{C}_1^{\prime\prime}(\nu)} \right\}
\end{equation}
and where $\widetilde{C}_1^{\prime\prime}(\nu)$ is obtained by interchanging $S_n^{\prime}$ for $S_n^{\prime\prime}$ in \eqref{eq:Ctilde1primeSC} above
and use $C^{\prime\prime}(0)=\sum_n S_n^{\prime\prime}$. The parameter $v_\mrm{s}^\prime$ is then
calculated based on the parameter $S_n^{\prime\prime}$ instead of $S_n^{\prime}$, similarly as in \eqref{eq:vssol2nu}.

Finally, the modified projection approach, here abbreviated as SC-mod, is obtained by replacing the function 
$\widetilde{C}_1^\prime(\nu)$ in \eqref{eq:Ctilde1primeSC} for
\begin{equation}\label{eq:Ctilde1primeSCmod}
\widetilde{C}_1^\prime(\nu)
=\sum_n \frac{S_n^\prime}{\gamma_{0n}^\prime+v_\mrm{s}^\prime\frac{S_n^\prime}{C^\prime(0)}-\iu\left(\nu-\nu_{0n}-\delta_{0n}^\prime\right)},
\end{equation}
and calculate $\sigma_\mrm{a}(\nu)$ as in \eqref{eq:sigmaaSC}. Here, the parameter $v_\mrm{s}^{\prime}$ is furthermore given by
\begin{equation}\label{eq:vssol1nu}
v_\mrm{s}^{\prime}=\frac{\displaystyle\sum_n \gamma_{0n}^\prime S_n^\prime \left(1-\frac{S_n^\prime}{C^\prime(0)} \right)}
{\displaystyle\sum_n S_n^\prime\left(1-\frac{S_n^\prime}{C^\prime(0)} \right)^2},
\end{equation}
for improved accuracy.

More rigorously, we may also incorporate the fluctuation-dissipation theorem, here abbreviated as SC-mod (fd),  
by calculating $\sigma_\mrm{a}(\nu)$ as in \eqref{eq:sigmaaSCfd}
where $\widetilde{C}_1^{\prime\prime}(\nu)$ is obtained by interchanging $S_n^{\prime}$ for $S_n^{\prime\prime}$ in \eqref{eq:Ctilde1primeSCmod} above
and use $C^{\prime\prime}(0)=\sum_n S_n^{\prime\prime}$ instead of $C^{\prime}(0)$. The parameter $v_\mrm{s}^\prime$ is then
calculated based on the parameter $S_n^{\prime\prime}$ instead of $S_n^{\prime}$, similarly as in \eqref{eq:vssol1nu}.
The modified projection approach is a technique that combines the high accuracy of the Lorentzian close to the line centers 
while at the same time maintaining the high accuracy of SC method in the far wings \cite{Nordebo2023a},
\cf in particular \cite[Eqs.~(64), (69), (70) and (74)]{Nordebo2023a}.

It is finally noticed that similar scalings taking the fluctuation-dissipation theorem (FDT) into account can readily be made based on
first order Rosenkranz parameters as explained in \cite{Nordebo2023a}. Other more advanced line mixing methods 
may also be investigated in the same way. However, the main emphasis here is to demonstrate the significance of using or not using the FDT in connection
with simple straightforward line mixing approaches, and one would certainly expect to see the same behavior with the more 
modern methods such as in \cite[Chapt.~4]{Hartmann+etal2021}.

\subsection{Numerical examples}

Two numerical examples are given below to illustrate the significance of the fluctuation-dissipation theorem in infrared spectroscopy.
The six different methods Lor, Lor (fd), SC, SC (fd), SC-mod and SC-mod (fd) defined in the previous section will be employed for this purpose.
The necessary spectroscopic parameters such as transition frequencies, line strengths and line widths and shifts, 
have all been retrieved from the HITRAN database \cite{Gordon+etal2022}.

\begin{figure}
\begin{center}
\includegraphics[width=0.48\textwidth]{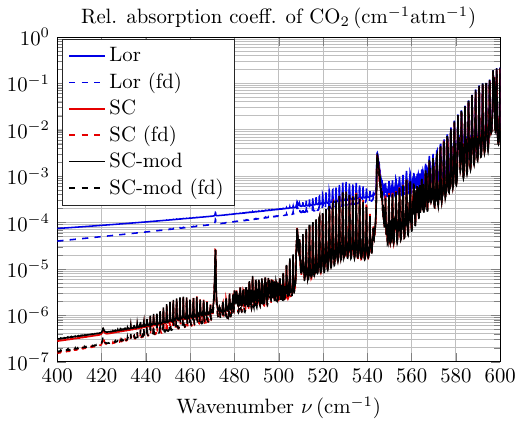}
\end{center}
\vspace{-5mm}
\caption{Relative absorption coefficient for the lower $\nu_2$ $\mrm{CO}_2$-band in dry air at $T=20$\unit{\degree C} and total pressure $p=1$\unit{atm}.
The blue, red and black lines indicate the sum of Lorentzian lines (Lor), the strong collision line mixing method (SC) and the modified SC-method (SC-mod), respectively.
The dashed and solid lines are showing computations with and without taking the fluctuation-dissipation (fd) theorem into account, respectively.
}
\label{fig:FluctDiss_fig4}
\end{figure}

In Fig.~\ref{fig:FluctDiss_fig4} is shown the first example which is concerned with the computation of the relative absorption coefficient for $\mrm{CO}_2$ in the lower end
of its $\nu_2$-band at $400$--$600$\unit{cm^{-1}}. 
The computations are for dry air at a temperature of $20$\unit{\degree C} and a total pressure of $1$\unit{atm}.
The line parameters are furthermore calculated for $1$\unit{\%} $\mrm{CO}_2$ and the absorption coefficient
is then scaled for path length in \unit{cm} and partial pressure in \unit{atm}.
We are furthermore employing here an external bandwidth $B_\mrm{ext}=300\unit{cm^{-1}}$
referring to the maximal distance between either endpoints of the spectral range (the computational domain)
and the transitions that are being included in the computations. This turned out to be more than enough to ensure convergence of the computed spectrum. 
All the available transitions which are inside the computational domain are always included.
In this first example, we have chosen to investigate the high-frequency end of the far infrared spectrum,
and we can confirm that the impact of not using the fluctuation-dissipation theorem (fd) is relatively small here.
It can also be confirmed here that there is only a very small deviation between the SC and the SC-mod methods,
which is expected since we are focusing here on low resolution far wing behavior rather than on the high resolution features close to the line centers, \cf \cite{Nordebo2023a}.
It is, however, very interesting to observe here the relatively large discrepancy between the simple sum of Lorentzian lines and the line mixing methods
which may have some impact on the calculation of atmospheric radiative transfer in this frequency range.

\begin{figure}
\begin{center}
\includegraphics[width=0.48\textwidth]{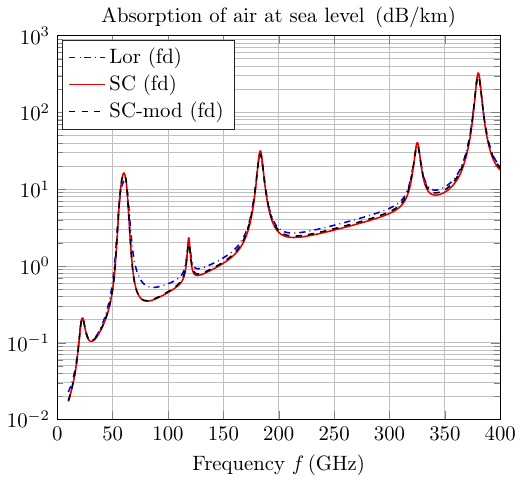}
\end{center}
\vspace{-5mm}
\caption{Absorption of moist air in the millimeter range at $15$\unit{\degree C}, total pressure $1$\unit{atm} and 60\unit{\%} humidity.
The blue dashdotted, red solid and black dashed lines indicate the sum of Lorentzian lines (Lor (fd)), the strong collision line mixing method (SC (fd)) 
and the modified SC-method (SC-mod (fd)), respectively, all of them taking the fluctuation-dissipation theorem into account.}
\label{fig:FluctDiss_fig6} 
\end{figure}

\begin{figure}
\begin{center}
\includegraphics[width=0.48\textwidth]{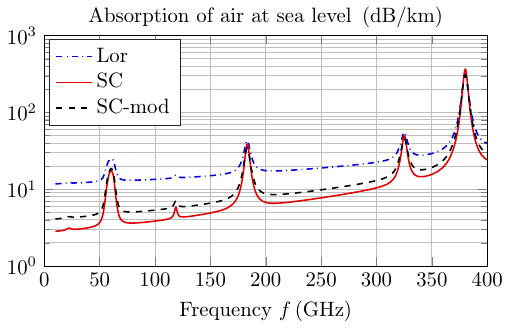}
\end{center}
\vspace{-5mm}
\caption{Absorption of moist air similar as in Fig.~\ref{fig:FluctDiss_fig6}.
The blue dashdotted, red solid and black dashed lines indicate the sum of Lorentzian lines (Lor), 
the strong collision line mixing method (SC) and the modified SC-method (SC-mod), respectively,
none of them taking the fluctuation-dissipation theorem into account.}
\label{fig:FluctDiss_fig7} 
\end{figure}

\begin{figure}
\begin{center}
\includegraphics[width=0.43\textwidth]{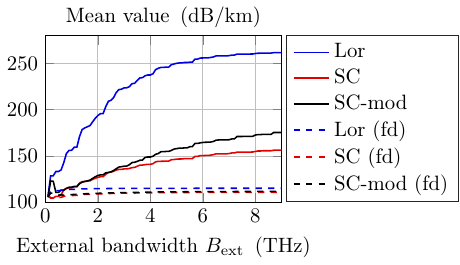}
\end{center}
\vspace{-5mm}
\caption{Mean value of absorption over the 400\unit{GHz} spectral range corresponding to the examples of Figs.~\ref{fig:FluctDiss_fig6} and \ref{fig:FluctDiss_fig7},
plotted here as a function of the bandwidth $B_\mrm{ext}$ of externally included transitions. }
\label{fig:FluctDiss_fig8} 
\end{figure}

The second example is in the low-frequency end of the far infrared spectrum.
Thus, we are considering here the atmospheric absorption of $\mrm{O}_2$ and $\mrm{H}_2\mrm{O}$ in the millimeter band up to 400\unit{GHz}.
The air is assumed to consist of 21\unit{\%} oxygen with a water vapor content corresponding to 60\unit{\%} humidity
at a temperature of $15$\unit{\degree C} and a total pressure of $1$\unit{atm}.
The results of the computations including the frequency dependency implied by the fluctuation-dissipation theorem (fd) are illustrated 
in Fig.~\ref{fig:FluctDiss_fig6}. These computations include also the water vapor continuum absorption as explained in 
connection with \cite[Fig.~4]{Nordebo2023a}, and the SC plots are almost identical with similar results shown in \cite[Fig.~4-6a on p.~124]{Richards+etal2010}.
Here, the external bandwidth is $B_\mrm{ext}=9$\unit{THz} ($300\unit{cm^{-1}}$) and which was more than enough to 
ensure convergence of the computed spectrum, \cf also Fig.~\ref{fig:FluctDiss_fig8}. 
Now, in this frequency range it is absolutely necessary to incorporate the frequency dependency 
implied by the fluctuation-dissipation theorem as expressed in \eqref{eq:sigmaaexprnu}.
As for a comparison, we can see in Fig.~\ref{fig:FluctDiss_fig7} the consequences of not taking the fluctuation-dissipation theorem into account.
Again, the external bandwidth was $B_\mrm{ext}=9$\unit{THz} ($300\unit{cm^{-1}}$) to secure the convergence of the computed spectrum.
Notice that these results are now clearly quite erroneus, except when close to the strongest line centers.
Moreover, it is also illustrated in Fig.~\ref{fig:FluctDiss_fig8} that the convergence of the computed spectrum with respect to the external bandwidth $B_\mrm{ext}$ 
becomes very slow when not incorporating the scaling \eqref{eq:Snbisdef} in accordance with the fluctuation-dissipation theorem.

\section{Summary}\label{sect:Summary}
A derivation of the fluctuation-dissipation theorem based on quantum electrodynamics (QED) has been presented in this paper.
The study is motivated by the fundamental importance of this widely cited semiclassical result whose detailed derivation and interpretation in terms of QED
seems to be very difficult to trace in the literature. Starting from the adequate statistical averages and by employing basic 
QED principles, it has been shown here that the theorem can be proved in a wider setting including the 
Schwarzschild's equation for radiative transfer. In particular, it is found that the classical and the QED interpretations differ in the following regards:
Firstly, in the classical interpretation the Beer-Bouguer-Lambert law is associated with absorption only, whereas the QED interpretation involves absorption as well
as stimulated emission via the fluctuation-dissipation theorem. Secondly, the source term of the Schwarzschild's equation which is understood classically in terms
of Kirchhoff's law of thermal radiation, can now be shown to be due solely to spontaneous emission independent of the applied radiation intensity.
The two interpretations are equivalent provided that the fluctuation-dissipation theorem is appropriately incorporated.
The significance of the fluctuation-dissipation theorem has furthermore been adressed 
in connection with far infrared and millimeter wave applications. In particular, it has been shown how an appropriate scaling of the line strength parameters including line mixing
can be carried out in the wavenumber domain. Numerical examples are given for the absorption of carbon dioxide in the lower end of its $\nu_2$ band
as well as for the atmospheric absorption of water vapor and oxygen for millimeter  waves up to 400\unit{GHz}.

\appendix
\subsection{Removed power}\label{sect:abspower}

\begin{figure}
\begin{center}
\includegraphics[width=0.50\textwidth]{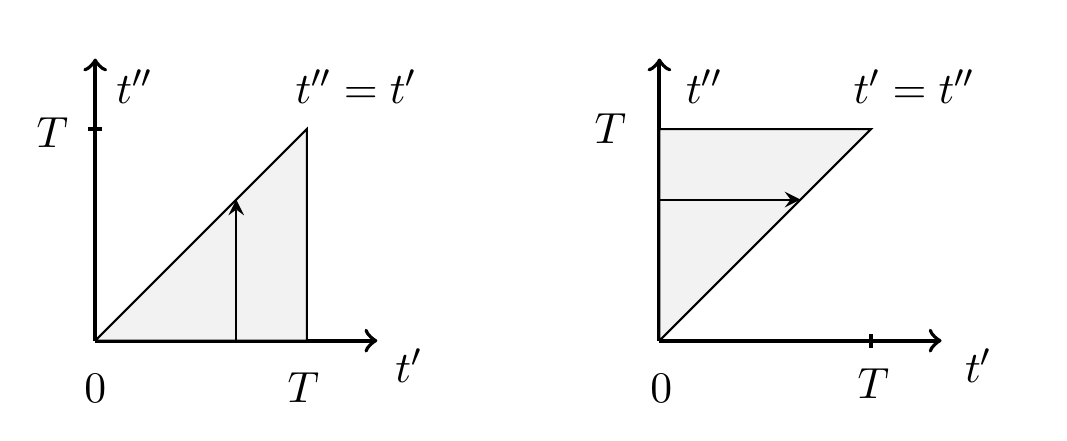}
\end{center}
\vspace{-5mm}
\caption{The two complementary triangular domains corresponding to the first two double integrals in \eqref{eq:doubleintdomains}, respectively.
The arrows indicate the direction of integration.
}
\label{fig:FDTfig1}
\end{figure}

We will now calculate the power that is removed from the radiation as defined in \eqref{eq:Padef}.
We start by investigating the expression
\begin{multline}
\trace\{\rho(0)H_{\bm{k}}\} - \trace\{\rho(T)H_{\bm{k}}\} \\
=\trace\left\{\rho(0)H_{\bm{k}} - \eu^{-\iu H_0T/\hbar}U_\mrm{I}(T)\rho(0)U_\mrm{I}^\dagger(T)\eu^{\iu H_0T/\hbar} H_{\bm{k}} \right\} \\
=\trace\left\{\rho(0)H_{\bm{k}} - U_\mrm{I}(T)\rho(0)U_\mrm{I}^\dagger(T) H_{\bm{k}}\right\},
\end{multline}
which is based on \eqref{eq:rhotsol} and the fact that $\eu^{\iu H_0T/\hbar}$ and $H_{\bm{k}}$ are commuting operators. 
And then by invoking the second order approximation in terms of the interaction potential,  we can insert
\eqref{eq:UI2ndorder} to obtain
\begin{multline}
\trace\{\rho(0)H_{\bm{k}}\} - \trace\{\rho(T)H_{\bm{k}}\} \\
=\trace\left\{\rho(0)H_{\bm{k}} -\left(1-\frac{\iu}{\hbar}\int_0^T V_\mrm{I}(t^\prime)\mrm{d}t^\prime \right.\right.\\
\left.\left. -\frac{1}{\hbar^2} \int_0^T \int_0^{t^\prime}V_\mrm{I}(t^\prime)V_\mrm{I}(t^{\prime\prime})\mrm{d}t^{\prime\prime}\mrm{d}t^\prime\right) \right. \\
\left. \cdot\ \rho(0)\left(1-\frac{\iu}{\hbar}\int_0^T V_\mrm{I}(t^\prime)\mrm{d}t^\prime \right. \right. \\
\left. \left. -\frac{1}{\hbar^2} \int_0^T \int_0^{t^\prime}V_\mrm{I}(t^\prime)V_\mrm{I}(t^{\prime\prime})\mrm{d}t^{\prime\prime}\mrm{d}t^\prime\right)^\dagger H_{\bm{k}}
\right\}.
\end{multline}
We can see immediately that the zero order contribution vanishes. Also the first order contribution vanishes since
\begin{multline}
\trace\left\{\int_0^T V_\mrm{I}(t^\prime)\mrm{d}t^\prime\rho(0) H_{\bm{k}}
- \rho(0)\int_0^T V_\mrm{I}^\dagger(t^\prime)\mrm{d}t^\prime H_{\bm{k}}\right\} \\
=\trace\left\{\int_0^T V_\mrm{I}(t^\prime)\mrm{d}t^\prime\rho(0) H_{\bm{k}}
- \int_0^T V_\mrm{I}(t^\prime)\mrm{d}t^\prime \rho(0) H_{\bm{k}}\right\} \\ =0,
\end{multline}
where we have exploited that $V_\mrm{I}^\dagger(t^\prime)=V_\mrm{I}(t^\prime)$ and $H_{\bm{k}}\rho(0)=\rho(0)H_{\bm{k}}$.
The remaining quadratic expression is now
\begin{multline}
\trace\{\rho(0)H_{\bm{k}}\} - \trace\{\rho(T)H_{\bm{k}}\} \\
=\trace\left\{ \frac{1}{\hbar^2}\int_0^T \int_0^{t^\prime}V_\mrm{I}(t^\prime)V_\mrm{I}(t^{\prime\prime})\mrm{d}t^{\prime\prime}\mrm{d}t^\prime\rho(0)H_{\bm{k}}\right\} \\
+\trace\left\{\rho(0)\frac{1}{\hbar^2} \int_0^T \int_0^{t^\prime}V_\mrm{I}(t^{\prime\prime})V_\mrm{I}(t^\prime)\mrm{d}t^{\prime\prime}\mrm{d}t^\prime H_{\bm{k}} \right\} \\
+\trace\left\{\frac{\iu}{\hbar}\int_0^T V_\mrm{I}(t^\prime)\mrm{d}t^\prime\rho(0)\frac{\iu}{\hbar}\int_0^T V_\mrm{I}(t^\prime)\mrm{d}t^\prime H_{\bm{k}}\right\},
\end{multline}
and where we notice that the operators $V_\mrm{I}(t^\prime)$ and $V_\mrm{I}(t^{\prime\prime})$ come in opposite order in the second integral due to the adjoint operation $(\cdot)^\dagger$.
Notice, however, that the order of integration is the same in the first and the second double integrals above, \ie first an integration over $t^{\prime\prime}$ and then over $t^\prime$.
Now, by replacing $t^{\prime\prime}$ for $t^\prime$ in the second double integral and similarly modify the third term, the expression can be manipulated to read
\begin{multline}\label{eq:doubleintdomains}
\trace\{\rho(0)H_{\bm{k}}\} - \trace\{\rho(T)H_{\bm{k}}\} \\
= \frac{1}{\hbar^2}\int_0^T \int_0^{t^\prime}\trace\left\{V_\mrm{I}(t^\prime)V_\mrm{I}(t^{\prime\prime}) H_{\bm{k}} \rho(0)\right\}\mrm{d}t^{\prime\prime}\mrm{d}t^\prime \\
+\frac{1}{\hbar^2} \int_0^T \int_0^{t^{\prime\prime}}\trace\left\{V_\mrm{I}(t^{\prime})V_\mrm{I}(t^{\prime\prime}) H_{\bm{k}}\rho(0) \right\}\mrm{d}t^\prime\mrm{d}t^{\prime\prime} \\
-\frac{1}{\hbar^2}\int_0^T\int_0^T \trace\left\{V_\mrm{I}(t^{\prime\prime})\rho(0) V_\mrm{I}(t^{\prime}) H_{\bm{k}}\right\}\mrm{d}t^\prime\mrm{d}t^{\prime\prime}.
\end{multline}
It is now observed that the first two double integrals above are defined over complementary triangular domains in the $t^{\prime}-t^{\prime\prime}$ plane as illustrated in Fig.~\ref{fig:FDTfig1}, 
and hence that we can write
\begin{multline}\label{eq:Numexpr}
\trace\{\rho(0)H_{\bm{k}}\} - \trace\{\rho(T)H_{\bm{k}}\} \\
=\frac{1}{\hbar^2} \int_0^T \int_0^{T}\trace\left\{V_\mrm{I}(t^{\prime})V_\mrm{I}(t^{\prime\prime}) H_{\bm{k}}\rho(0) \right\}\mrm{d}t^\prime\mrm{d}t^{\prime\prime} \\
-\frac{1}{\hbar^2}\int_0^T\int_0^T \trace\left\{V_\mrm{I}(t^\prime) H_{\bm{k}} V_\mrm{I}(t^{\prime\prime}) \rho(0) \right\}\mrm{d}t^\prime\mrm{d}t^{\prime\prime},
\end{multline}
and where also the last term has been modified using the trace rule $\trace\{AB\}=\trace\{BA\}$.

Now, we recall that $V_\mrm{I}(t)=\eu^{\iu H_0t/\hbar}(-d{\cal E})\eu^{-\iu H_0t/\hbar}$ and $\rho(0)=\rho_{\bm{k}}\rho_\mrm{M}$, and consider the two terms
\begin{multline}
T_1=\frac{1}{\hbar^2} \int_0^T \int_0^{T}\trace\left\{
\eu^{\iu H_0t^{\prime}/\hbar}{\cal E}d\eu^{-\iu H_0t^{\prime}/\hbar} \right. \\
\left. \cdot\ \eu^{\iu H_0t^{\prime\prime}/\hbar}d{\cal E}\eu^{-\iu H_0t^{\prime\prime}/\hbar}
H_{\bm{k}}\rho_{\bm{k}}\rho_\mrm{M} \right\}\mrm{d}t^\prime\mrm{d}t^{\prime\prime}
\end{multline}
and
\begin{multline}
T_2=\frac{1}{\hbar^2} \int_0^T \int_0^{T}\trace\left\{
\eu^{\iu H_0t^{\prime}/\hbar}{\cal E}d\eu^{-\iu H_0t^{\prime}/\hbar} \right. \\
\left. \cdot\ H_{\bm{k}}\eu^{\iu H_0t^{\prime\prime}/\hbar}d{\cal E}\eu^{-\iu H_0t^{\prime\prime}/\hbar}
\rho_{\bm{k}}\rho_\mrm{M} \right\}\mrm{d}t^\prime\mrm{d}t^{\prime\prime}.
\end{multline}
By using $H_0=H_\mrm{M}+H_{\bm{k}}$, both terms are further expanded as
\begin{multline}
T_1=\frac{1}{\hbar^2} \int_0^T \int_0^{T} \\
\trace\left\{\eu^{\iu H_\mrm{M}t^{\prime}/\hbar}\eu^{\iu H_{\bm{k}}t^{\prime}/\hbar}{\cal E}d
\eu^{-\iu H_\mrm{M}(t^{\prime}-t^{\prime\prime})/\hbar}\eu^{-\iu H_{\bm{k}}(t^{\prime}-t^{\prime\prime})/\hbar}  \right. \\
\left. \cdot\ d{\cal E}\eu^{-\iu H_\mrm{M}t^{\prime\prime}/\hbar}\eu^{-\iu H_{\bm{k}}t^{\prime\prime}/\hbar}
H_{\bm{k}}\rho_{\bm{k}}\rho_\mrm{M} \right\}\mrm{d}t^\prime\mrm{d}t^{\prime\prime},
\end{multline}
and
\begin{multline}
T_2=\frac{1}{\hbar^2} \int_0^T \int_0^{T} \\
\trace\left\{
\eu^{\iu H_\mrm{M}t^{\prime}/\hbar}\eu^{\iu H_{\bm{k}}t^{\prime}/\hbar}{\cal E}d
\eu^{-\iu H_\mrm{M}(t^{\prime}-t^{\prime\prime})/\hbar}\eu^{-\iu H_{\bm{k}}(t^{\prime}-t^{\prime\prime})/\hbar}  \right. \\
\left. \cdot\ dH_{\bm{k}}{\cal E} \eu^{-\iu H_\mrm{M}t^{\prime\prime}/\hbar}\eu^{-\iu H_{\bm{k}}t^{\prime\prime}/\hbar}
\rho_{\bm{k}}\rho_\mrm{M} \right\}\mrm{d}t^\prime\mrm{d}t^{\prime\prime},
\end{multline}
and where we in the last line have exploited the fact that $H_{\bm{k}}$ commutes with $\eu^{\iu H_0t^{\prime\prime}/\hbar}$ as well as with $d$.

We can now take the partial trace of these expressions by applying $\bra{n}\cdot\ket{n}$, yielding
\begin{multline}\label{eq:T1traceM0}
T_1=\frac{1}{\hbar^2} \int_0^T \int_0^{T}\trace\left\{
\eu^{\iu H_\mrm{M}t^{\prime}/\hbar}\eu^{\iu (n+\frac{1}{2})\omega t^{\prime}} \right. \\
\left. \cdot\  \bra{n}{{\cal E}}d\eu^{-\iu H_\mrm{M}(t^{\prime}-t^{\prime\prime})/\hbar}\eu^{-\iu H_{\bm{k}}(t^{\prime}-t^{\prime\prime})/\hbar}
d{{\cal E}}\ket{n} \right. \\
\left. \cdot\  \eu^{-\iu H_\mrm{M}t^{\prime\prime}/\hbar}\eu^{-\iu (n+\frac{1}{2})\omega t^{\prime\prime}}
(n+\frac{1}{2})\hbar\omega w_n\rho_\mrm{M} \right\}\mrm{d}t^\prime\mrm{d}t^{\prime\prime},
\end{multline}
and
\begin{multline}\label{eq:T2traceM0}
T_2=\frac{1}{\hbar^2} \int_0^T \int_0^{T}\trace\left\{
\eu^{\iu H_\mrm{M}t^{\prime}/\hbar}\eu^{\iu (n+\frac{1}{2})\omega t^{\prime}} \right. \\
\left. \cdot\  \bra{n}{{\cal E}}d \eu^{-\iu H_\mrm{M}(t^{\prime}-t^{\prime\prime})/\hbar}\eu^{-\iu H_{\bm{k}}(t^{\prime}-t^{\prime\prime})/\hbar}
dH_{\bm{k}}{{\cal E}}\ket{n} \right. \\
\left. \cdot\  \eu^{-\iu H_\mrm{M}t^{\prime\prime}/\hbar}\eu^{-\iu (n+\frac{1}{2})\omega t^{\prime\prime}}
w_n \rho_\mrm{M} \right\}\mrm{d}t^\prime\mrm{d}t^{\prime\prime},
\end{multline}
and where the remaining trace is taken with respect to the total molecular system without radiation.

We will now investigate the partial traces $\bra{n}\cdot\ket{n}$ above. We start by reviewing \eqref{eq:EB} and \eqref{eq:anahatn} and notice that
\begin{equation}\label{eq:EnandnE}
\left\{\begin{array}{l}
{{\cal E}}\ket{n}={{\cal E}}_0\iu\omega\left(\sqrt{n}\ket{n-1}\eu^{\iu\bm{k}\cdot\bm{r}}-\sqrt{n+1}\ket{n+1}\eu^{-\iu\bm{k}\cdot\bm{r}} \right), \vspace{0.2cm} \\
\bra{n}{{\cal E}}=-{{\cal E}}_0\iu\omega\left(\sqrt{n}\bra{n-1}\eu^{-\iu\bm{k}\cdot\bm{r}}-\sqrt{n+1}\bra{n+1}\eu^{\iu\bm{k}\cdot\bm{r}} \right).
\end{array}\right.
\end{equation}
We can now evaluate
\begin{multline}
\bra{n}{{\cal E}}d\eu^{-\iu H_\mrm{M}(t^{\prime}-t^{\prime\prime})/\hbar}\eu^{-\iu H_{\bm{k}}(t^{\prime}-t^{\prime\prime})/\hbar}d{{\cal E}}\ket{n} \\
=-{{\cal E}}_0\iu\omega\left(\sqrt{n}\bra{n-1}\eu^{-\iu\bm{k}\cdot\bm{r}}-\sqrt{n+1}\bra{n+1}\eu^{\iu\bm{k}\cdot\bm{r}} \right) \\
\cdot\  d\eu^{-\iu H_\mrm{M}(t^{\prime}-t^{\prime\prime})/\hbar}d {{\cal E}}_0\iu\omega\\
\cdot\  \left(\sqrt{n}\ket{n-1}\eu^{\iu\bm{k}\cdot\bm{r}}\eu^{-\iu(n-1+\frac{1}{2})\omega(t^{\prime}-t^{\prime\prime})}  \right. \\
\left. -\sqrt{n+1}\ket{n+1}\eu^{-\iu\bm{k}\cdot\bm{r}}\eu^{-\iu(n+1+\frac{1}{2})\omega(t^{\prime}-t^{\prime\prime})} \right) \\
={{\cal E}}_0^2\omega^2\left\{ n\eu^{-\iu\bm{k}\cdot\bm{r}}d\eu^{-\iu H_\mrm{M}(t^{\prime}-t^{\prime\prime})/\hbar}d\eu^{\iu\bm{k}\cdot\bm{r}}\eu^{-\iu(n-1+\frac{1}{2})\omega(t^{\prime}-t^{\prime\prime})} \right. \\
\left. +(n+1)\eu^{\iu\bm{k}\cdot\bm{r}}d\eu^{-\iu H_\mrm{M}(t^{\prime}-t^{\prime\prime})/\hbar}d\eu^{-\iu\bm{k}\cdot\bm{r}}\eu^{-\iu(n+1+\frac{1}{2})\omega(t^{\prime}-t^{\prime\prime})} 
\right\},
\end{multline}
and
\begin{multline}
\bra{n}{{\cal E}}d\eu^{-\iu H_\mrm{M}(t^{\prime}-t^{\prime\prime})/\hbar}\eu^{-\iu H_{\bm{k}}(t^{\prime}-t^{\prime\prime})/\hbar}dH_{\bm{k}}{{\cal E}}\ket{n} \\
=-{{\cal E}}_0\iu\omega\left(\sqrt{n}\bra{n-1}\eu^{-\iu\bm{k}\cdot\bm{r}}-\sqrt{n+1}\bra{n+1}\eu^{\iu\bm{k}\cdot\bm{r}} \right) \\
\cdot\ d\eu^{-\iu H_\mrm{M}(t^{\prime}-t^{\prime\prime})/\hbar}d {{\cal E}}_0\iu\omega \\
\cdot\  \left(\sqrt{n}\ket{n-1}\eu^{\iu\bm{k}\cdot\bm{r}}\eu^{-\iu(n-1+\frac{1}{2})\omega(t^{\prime}-t^{\prime\prime})} (n-1+\frac{1}{2})\hbar\omega \right. \\
\left. -\sqrt{n+1}\ket{n+1}\eu^{-\iu\bm{k}\cdot\bm{r}}\eu^{-\iu(n+1+\frac{1}{2})\omega(t^{\prime}-t^{\prime\prime})} (n+1+\frac{1}{2})\hbar\omega \right) \\
={{\cal E}}_0^2\omega^2\left\{ n\eu^{-\iu\bm{k}\cdot\bm{r}}d\eu^{-\iu H_\mrm{M}(t^{\prime}-t^{\prime\prime})/\hbar} 
d\eu^{\iu\bm{k}\cdot\bm{r}}\eu^{-\iu(n-1+\frac{1}{2})\omega(t^{\prime}-t^{\prime\prime})} \right. \\
\left. \cdot\  
(n-1+\frac{1}{2})\hbar\omega \right. \\
\left. +(n+1)\eu^{\iu\bm{k}\cdot\bm{r}}d\eu^{-\iu H_\mrm{M}(t^{\prime}-t^{\prime\prime})/\hbar}d\eu^{-\iu\bm{k}\cdot\bm{r}}\eu^{-\iu(n+1+\frac{1}{2})\omega(t^{\prime}-t^{\prime\prime})}  \right. \\
\left. \cdot\  (n+1+\frac{1}{2})\hbar\omega \right\}.
\end{multline}

We can now insert these results into \eqref{eq:T1traceM0} and \eqref{eq:T2traceM0} and simplify, to get
\begin{multline}\label{eq:T1traceM1}
T_1=\frac{{{\cal E}}_0^2\omega^2}{\hbar^2} \int_0^T \int_0^{T}\trace\left\{
\eu^{\iu H_\mrm{M}(t^{\prime}-t^{\prime\prime})/\hbar} \right. \\
\left. \cdot\  \left(n\eu^{-\iu\bm{k}\cdot\bm{r}}d\eu^{-\iu H_\mrm{M}(t^{\prime}-t^{\prime\prime})/\hbar}d\eu^{\iu\bm{k}\cdot\bm{r}}\eu^{\iu\omega(t^{\prime}-t^{\prime\prime})} \right.\right. \\
\left.\left. +(n+1)\eu^{\iu\bm{k}\cdot\bm{r}}d\eu^{-\iu H_\mrm{M}(t^{\prime}-t^{\prime\prime})/\hbar}d\eu^{-\iu\bm{k}\cdot\bm{r}}\eu^{-\iu\omega(t^{\prime}-t^{\prime\prime})} 
\right) \right. \\
\left. \cdot\  (n+\frac{1}{2})\hbar\omega w_n\rho_\mrm{M} \right\}\mrm{d}t^\prime\mrm{d}t^{\prime\prime},
\end{multline}
and
\begin{multline}\label{eq:T2traceM1}
T_2=\frac{{{\cal E}}_0^2\omega^2}{\hbar^2} \int_0^T \int_0^{T}\trace\left\{
\eu^{\iu H_\mrm{M}(t^{\prime}-t^{\prime\prime})/\hbar} \right. \\
\left. \cdot\  \left(n\eu^{-\iu\bm{k}\cdot\bm{r}}d\eu^{-\iu H_\mrm{M}(t^{\prime}-t^{\prime\prime})/\hbar}d\eu^{\iu\bm{k}\cdot\bm{r}}\eu^{\iu\omega(t^{\prime}-t^{\prime\prime})}
(n-1+\frac{1}{2})\hbar\omega \right.\right. \\
 +(n+1)\eu^{\iu\bm{k}\cdot\bm{r}}d\eu^{-\iu H_\mrm{M}(t^{\prime}-t^{\prime\prime})/\hbar}d\eu^{-\iu\bm{k}\cdot\bm{r}}\eu^{-\iu\omega(t^{\prime}-t^{\prime\prime})}  \\
\left. \left. \cdot\  (n+1+\frac{1}{2})\hbar\omega \right) w_n \rho_\mrm{M} \right\}\mrm{d}t^\prime\mrm{d}t^{\prime\prime}.
\end{multline}

The results above can now be summarized as
\begin{multline}\label{eq:Numexpr2}
\trace\{\rho(0)H_{\bm{k}}\} - \trace\{\rho(T)H_{\bm{k}}\}=T_1-T_2 \\
=\frac{{{\cal E}}_0^2\omega^3}{\hbar} \int_0^T \int_0^{T}\mrm{d}t^\prime\mrm{d}t^{\prime\prime}\left\{ \right. \\
\trace\left\{\eu^{\iu H_\mrm{M}(t^{\prime}-t^{\prime\prime})/\hbar}d\eu^{-\iu\bm{k}\cdot\bm{r}}\eu^{-\iu H_\mrm{M}(t^{\prime}-t^{\prime\prime})/\hbar}d\eu^{\iu\bm{k}\cdot\bm{r}}
\rho_\mrm{M} \right\} \\
\cdot\  \eu^{\iu\omega(t^{\prime}-t^{\prime\prime})} nw_n \\
-\trace\left\{\eu^{\iu H_\mrm{M}(t^{\prime}-t^{\prime\prime})/\hbar}d\eu^{\iu\bm{k}\cdot\bm{r}}\eu^{-\iu H_\mrm{M}(t^{\prime}-t^{\prime\prime})/\hbar}d\eu^{-\iu\bm{k}\cdot\bm{r}}
\rho_\mrm{M} \right\} \\
\left. \cdot\  \eu^{-\iu\omega(t^{\prime}-t^{\prime\prime})}(n+1)w_n\right\}. 
\end{multline}

\begin{figure}
\begin{center}
\includegraphics[width=0.50\textwidth]{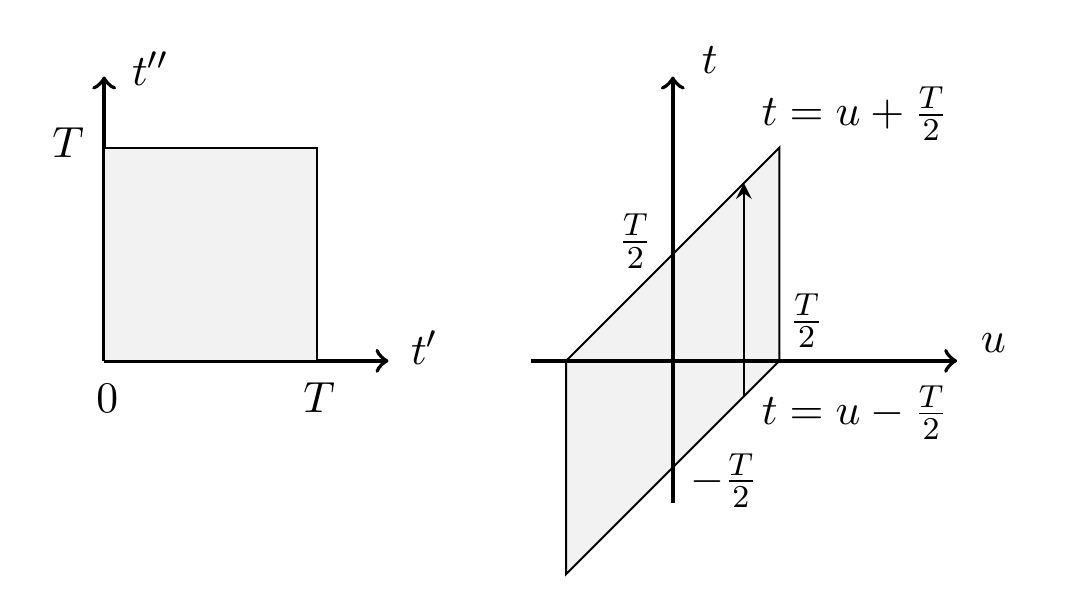}
\end{center}
\vspace{-5mm}
\caption{Graphical illustration of the mapping $(t^\prime,t^{\prime\prime})\rightarrow (u,t)$ defined in \eqref{eq:tumapping}.
The arrow in the right figure indicates the direction of integration in the $t$-domain.
}
\label{fig:FDTfig2}
\end{figure}

Let us now define the dipole autocorrelation function
\begin{equation}\label{eq:Cktexpr}
C(\bm{k},t)=\trace\left\{
\eu^{\iu H_\mrm{M}t/\hbar}d\eu^{-\iu\bm{k}\cdot\bm{r}}\eu^{-\iu H_\mrm{M}t/\hbar}d\eu^{\iu\bm{k}\cdot\bm{r}}\rho_\mrm{M} \right\},
\end{equation}
and its Fourier transform
\begin{equation}
I(\bm{k},\omega)=\frac{1}{2\pi}\int_{-\infty}^{\infty}C(\bm{k},t)\eu^{\iu\omega t}\mrm{d}t,
\end{equation}
similarly as in \eqref{eq:QMCtsummary1} and \eqref{eq:IomegaFT0}.
It is also observed that the integrands in the double integrals in \eqref{eq:Numexpr2} depend solely on the time-difference $t^{\prime}-t^{\prime\prime}$.
We can then introduce the variable substitution
\begin{equation}\label{eq:tumapping}
\left\{\begin{array}{l}
u=t^{\prime}-\frac{T}{2} \vspace{0.2cm} \\
t= t^{\prime}-t^{\prime\prime},
\end{array}\right.
\end{equation}
as illustrated in Fig.~\ref{fig:FDTfig2} and where $\mrm{d}t^\prime\mrm{d}t^{\prime\prime}=\mrm{d}u\mrm{d}t$.
The integrals above then transform as
\begin{multline}\label{eq:intlimits}
\frac{1}{T}\int_0^T \int_0^{T} f(t^{\prime}-t^{\prime\prime})\mrm{d}t^\prime\mrm{d}t^{\prime\prime} \\
=\frac{1}{T}\int_{-T/2}^{T/2} \int_{u-T/2}^{u+T/2} f(t)\mrm{d}t\mrm{d}u\rightarrow \int_{-\infty}^\infty f(t)\mrm{d}t,
\end{multline}
where $f(t)$ is either integrand in \eqref{eq:Numexpr2} and where the limit is taken as $T\rightarrow \infty$, assuming that the last integral exists.

Based on the results given in \eqref{eq:Numexpr2} and \eqref{eq:intlimits} above we can now see that
\begin{multline}\label{eq:Pares3}
P_\mrm{r}=\displaystyle\lim_{T\rightarrow\infty}\frac{1}{T}\left(\trace\{\rho(0)H_{\bm{k}}\} - \trace\{\rho(T)H_{\bm{k}}\}\right) \\
=\frac{2\pi{\cal E}_0^2\omega^3}{\hbar} \left(I(\bm{k},\omega)\langle n \rangle- I(-\bm{k},-\omega)(\langle n \rangle +1)\right).
\end{multline}
Notice that the partial trace taken over the number states $\ket{n}$ has been expressed in \eqref{eq:Numexpr2}
using Einsteins summation convention and that the mean number of photons $\langle n \rangle$ is hence given by \eqref{eq:sumwnn}.

The final result as stated in \eqref{eq:Pares1} is obtained by redefining the dipole autocorrelation function \eqref{eq:QMCtsummary1} and \eqref{eq:Cktexpr}
to encompass a vector valued dipole operator $\bm{d}$, extending the trace operator accordingly and thus divide the expression in \eqref{eq:Pares3} by 3.
In this way, the vector nature of light is taken into account for isotropic conditions. Details are given in Appendix \ref{sect:vector} below.

\subsection{The vector nature of light}\label{sect:vector}

Let us now finally comment very briefly on the vector nature of light and its implications for the results described in this paper.
A more complete description of the electromagnetic field operator including its polarization is to write
\begin{equation}\label{eq:Evector}
\bm{{\cal E}}={\cal E}_0\iu\omega\left(\hat{\bm{e}}\hat{a}\eu^{\iu \bm{k}\cdot\bm{r}}-\hat{\bm{e}}^*\hat{a}^\dagger\eu^{-\iu \bm{k}\cdot\bm{r}} \right),
\end{equation}
where $\hat{\bm{e}}$ is a real or complex valued unit vector, \cf \eg \cite[p.~319]{Grynberg+etal2010} and \cite[p.~141]{Loudon2000}.
With regard to the interaction potential treated above the quantity $d{\cal E}$ is now replaced by
\begin{equation}
\bm{d}\cdot \bm{{\cal E}}={\cal E}_0\iu\omega\left(\hat{a}(\bm{d}\cdot\hat{\bm{e}})\eu^{\iu \bm{k}\cdot\bm{r}}
-\hat{a}^\dagger (\bm{d}\cdot\hat{\bm{e}}^*)\eu^{-\iu \bm{k}\cdot\bm{r}} \right),
\end{equation}
where $\bm{d}$ is the vector valued dipole operator. 
By comparison with \eqref{eq:EB}, we see that the following replacements should be implemented in all of the computations that have been carried out above
\begin{equation}\label{eq:replacements}
\left\{\begin{array}{l}
d\eu^{\iu \bm{k}\cdot\bm{r}}\rightarrow (\bm{d}\cdot\hat{\bm{e}})\eu^{\iu \bm{k}\cdot\bm{r}}, \vspace{0.2cm} \\
d\eu^{-\iu \bm{k}\cdot\bm{r}}\rightarrow (\bm{d}\cdot\hat{\bm{e}}^*)\eu^{-\iu \bm{k}\cdot\bm{r}}.
\end{array}\right.
\end{equation}
This means that the expressions \eqref{eq:QMCtsummary1} and \eqref{eq:Cktexpr} should be interpreted with respect to a particular polarization $\hat{\bm{e}}$ as
\begin{multline}\label{eq:Cktexpr2}
C_{\hat{\mrm{e}}}(\bm{k},t) \\
=\trace\left\{\eu^{\iu H_\mrm{M}t/\hbar}(\bm{d}\cdot\hat{\bm{e}}^*)\eu^{-\iu\bm{k}\cdot\bm{r}}\eu^{-\iu H_\mrm{M}t/\hbar}(\bm{d}\cdot\hat{\bm{e}})\eu^{\iu\bm{k}\cdot\bm{r}}\rho_\mrm{M} \right\} \\
=\hat{\bm{e}}^*\cdot\trace\left\{\eu^{\iu H_\mrm{M}t/\hbar}\bm{d}\eu^{-\iu\bm{k}\cdot\bm{r}}\eu^{-\iu H_\mrm{M}t/\hbar}\bm{d}\eu^{\iu\bm{k}\cdot\bm{r}}\rho_\mrm{M} \right\}\cdot\hat{\bm{e}},
\end{multline}
and where the trace is taken only over the quantum mechanical states that are associated with the Hamiltonian $H_\mrm{M}$.
Let us now redefine the dipole autocorrelation function as
\begin{equation}\label{eq:Cktexpr3}
C(\bm{k},t)=\trace\left\{\eu^{\iu H_\mrm{M}t/\hbar}\bm{d}\eu^{-\iu\bm{k}\cdot\bm{r}}\eu^{-\iu H_\mrm{M}t/\hbar}\bm{d}\eu^{\iu\bm{k}\cdot\bm{r}}\rho_\mrm{M} \right\},
\end{equation}
in accordance with \cite[Eq.~(II.8)]{Hartmann+etal2021} and where the trace also incorporates the three-dimensional space
spanned by the orthonormal cartesian vectors $\hat{\bm{e}}_1$, $\hat{\bm{e}}_2$, $\hat{\bm{e}}_3$.
By taking the mean value of \eqref{eq:Cktexpr2} over all polarization directions, we obtain
\begin{equation}
\langle C_{\hat{\mrm{e}}}(\bm{k},t) \rangle = \frac{1}{3}\sum_{i=1}^3 C_{\hat{\mrm{e}}_i}(\bm{k},t) = \frac{1}{3}C(\bm{k},t),
\end{equation}
and which explains the division by 3 that has been included in connection with \eqref{eq:sigmaaexprGauss}, \eqref{eq:Pares}, \eqref{eq:sigmaaexpr} and \eqref{eq:Pares1}.



\end{document}